%% file: science_template.tex
\documentclass[12pt]{article}

\usepackage{newtxtext,newtxmath}

\usepackage{graphicx}

\usepackage[letterpaper,margin=1in]{geometry}
\usepackage[normalem]{ulem}

\renewenvironment{abstract}
	{\quotation}
	{\endquotation}

\date{}

\makeatletter
\renewcommand{\fnum@figure}{\textbf{Figure \thefigure}}
\renewcommand{\fnum@table}{\textbf{Table \thetable}}
\makeatother

\usepackage{scicite}
\usepackage{lineno}

\usepackage{url}

\usepackage{amsmath}
\usepackage{graphicx}%
\usepackage{multirow}%
\usepackage{mathrsfs}%
\usepackage[title]{appendix}%
\usepackage{xcolor}%
\usepackage{textcomp}%
\usepackage{manyfoot}%
\usepackage{booktabs}%
\usepackage{algorithm}%
\usepackage{algorithmicx}%
\usepackage{algpseudocode}%
\usepackage{listings}%
\usepackage{caption}

\definecolor{fuchsia}{rgb}{0.54, 0.17, 0.89}
\definecolor{azure}{rgb}{0.0, 0.5, 1.0}
\definecolor{pgreen}{rgb}{0.12, 0.3, 0.17}
\definecolor{alizarin}{rgb}{0.82, 0.1, 0.26}

\newcommand{\oiii}{[\textrm{O}~\textsc{iii}]}
\newcommand{\oiiir}{[\textrm{O}~\textsc{iii}]_{\lambda5008}}

\newcommand{\oiiib}{[\textrm{O}~\textsc{iii}]_{\lambda4960}}
\newcommand{\oiiif}{$[\textrm{O}~\textsc{iii}]_{\lambda4363}$}

\newcommand{\simgt}{\,\rlap{\lower 3.5 pt \hbox{$\mathchar \sim$}} \raise
1pt \hbox {$>$}\,}
\newcommand{\simlt}{\,\rlap{\lower 3.5 pt \hbox{$\mathchar \sim$}} \raise
1pt \hbox {$<$}\,}

\newcommand{\logoh}{$\log {\rm (O/H)}$}

\newcommand{\ly}{${\rm Ly\alpha}$}
\newcommand{\ha}{${\rm H\alpha}$}
\newcommand{\hb}{${\rm H\beta}$}

\newcommand{\id}{AMORE6} 
\newcommand{\ida}{\id-A}
\newcommand{\idb}{\id-B}
\newcommand{\zspec}{5.725}
\newcommand{\zspecerr}{$5.7253_{-0.0005}^{+0.0005}$}
\newcommand{\qsigma}{3} 

\newcommand{\oxygensand}{6.22} 
\newcommand{\oxygensunsand}{$0.34\,\%$}

\newcommand{\oxygentwosigma}{5.79} 
\newcommand{\oxygen}{5.97} 
\newcommand{\oxygensuntwosigma}{$0.12\,\%$} 
\newcommand{\oxygensun}{$0.19\,\%$} 

\newcommand{\magnif}{78}
\newcommand{\magniferr}{$\magnif_{-6}^{+8}$}

\newcommand{\magnifaerr}{$39.3_{-3.5}^{+3.7}$}

\newcommand{\magnifberr}{$77.7_{-5.9}^{+8.7}$}

\newcommand{\magnifbfroma}{$55$}

\newcommand{\mstele}{$4.4_{-1.1}^{+2.7}\times10^5\,M_\odot$}
\newcommand{\mstelemagnif}{$4.4_{-1.1}^{+2.7}\times10^5\times(78/\mu)\,M_\odot$}
\newcommand{\sfremagnif}{$0.07_{-0.01}^{+0.01}\times(78/\mu)\,M_\odot {\rm yr^{-1}}$}
\newcommand{\uvbeta}{$-2.77_{-0.09}^{+0.07}$}
\newcommand{\uvbetaphot}{$-2.93_{-0.46}^{+0.41}$}
\newcommand{\uvbetastel}{$-3.00_{-0.01}^{+0.01}$}
\newcommand{\sSFR}{$149_{-24}^{+24}$}
\newcommand{\muv}{$-14.5_{-0.1}^{+0.1}$\,ABmag}

\newcommand{\reamas}{$39.5_{-3.1}^{+3.1}$\,mas} 
\newcommand{\rebmas}{$47.6_{-1.9}^{+1.9}$\,mas} 
\newcommand{\rea}{$36.9_{-4.4}^{+4.8}$\,pc} 
\newcommand{\reb}{$31.6_{-2.8}^{+2.6}$\,pc} 
\newcommand{\rebfroma}{$37.6_{-3.3}^{+3.1}$\,pc} 
\newcommand{\re}{$34.2_{-3.6}^{+3.7}$\,pc} 

\newcommand{\relenst}{$36.0_{-7.5}^{+9.4}$\,pc} 
\newcommand{\realenst}{$39.1_{-6.6}^{+7.9}$\,pc} 
\newcommand{\reblenst}{$32.9_{-3.6}^{+5.0}$\,pc} 

\newcommand{\lytohb}{$14.7\pm2.0$} 
\newcommand{\lyesc}{$63_{-9}^{+8}\,\%$} 
\newcommand{\snra}{4.2} 
\newcommand{\snrb}{6.2}
\newcommand{\snr}{8.2}
\newcommand{\SIGMAS}{$56.8_{-1.8}^{+2.0}\,M_\odot$\,pc$^{-2}$}
\newcommand{\SIGSFR}{$8.5_{-1.7}^{+1.6}\,M_\odot$\,yr$^{-1}$\,kpc$^{-2}$}

\newcommand{\ewha}{EW$_0({\rm H}\alpha)$} %
\newcommand{\ewhb}{EW$_0({\rm H}\beta)$} %
\newcommand{\ewhberr}{$940\pm150$\,\AA} 
\newcommand{\ewhbphotom}{$721_{-612}^{+1783}$\,\AA} 
\newcommand{\ewhaphotom}{$1910_{-1111}^{+3203}$\,\AA}
\newcommand{\dvhbly}{$14_{-150}^{+150}$\,km\,s$^{-1}$}%
\newcommand{\tage}{$1.5_{-1.0}^{+1.2}$\,Myr}

\usepackage[T1]{fontenc}

\def\scititle{
	Pristine Massive Star Formation Caught at the Break of Cosmic Dawn
}
\title{\bfseries \boldmath \scititle}

\input{authors}

\begin{document} 

\maketitle

\begin{abstract} \bfseries \boldmath
{
The existence of galaxies with no heavy elements is a key prediction of cosmological models. So far no ``zero-metallicity'', or Population~III, galaxies have been identified. Here, we report the identification of an extremely metal-poor galaxy ``\id'' at redshift $z=$\,\zspecerr, multiply imaged by a foreground galaxy cluster. JWST spectra consistently detect ${\rm H\beta}$ at both positions, but [\textrm{O}\textsc{iii}]$_{\lambda\lambda4960,5008}$ remains undetected. This places a firm upper limit on its oxygen abundance, {$<$\,\oxygensun\ of solar metallicity at \qsigma\,$\sigma$}, establishing itself as the most pristine galaxy by far. \id\ exhibits exceptional properties that indicate the presence of pristine massive star formation. Finding such an example at a relatively late cosmic time is surprising, but it also validates the basic ideas behind the Big Bang model.
}
\end{abstract}

\noindent
The existence of galaxies with no heavy elements -- elements heavier than Helium that were formed by stars after Big Bang nucleosynthesis -- is a key prediction of cosmological models. Finding them would provide direct and transformative evidence for the formation of the first galaxies in extremely different conditions than those observed today. The James Webb Space Telescope (JWST) has enabled the identification of galaxies all the way to cosmic times of a few hundred million years since it began scientific operations \cite{curtis-lake23,roberts-borsani23, bunkerGNz11,castellano24,carniani24,naidu25}. 
However, most galaxies observed so far exhibit strong emission lines originating in heavy elements, leaving ``zero-metallicity'' (or Population~III) galaxies yet to be discovered. The main conclusion of studies using statistical samples is that while the mass–metallicity relation evolves with redshift, massive galaxies were already significantly enriched within the first few hundred million years after the Big Bang \cite{nakajima23,laseter23,heintz23,sanders24,morishita24b,scholte25}. On the other hand, lower-metallicity galaxies have been identified at masses of $M_*<10^{8} M_\odot$ \cite{vanzella23,chemerynska24,hsiao25}, indicating that the evolution of metallicity may be driven more by galactic mass than by cosmic time.


\subsubsection*{Observations and Analysis}
The doubly lensed system 
``33.1a/b''
was first observed with VLT/MUSE \cite{mahler18,bergamini23}.
The intrinsic source, located behind the galaxy cluster  Abell~2744, is strongly gravitationally lensed, displaying two images to the west of the foreground cluster, {with magnification factors of $\mu=$\,\magnifaerr\ and \magnifberr, respectively \cite{bergamini23b}}. \ly\ emission was detected at both positions in the imaging plane, firmly establishing its redshift as $z=\zspec$. The same field has been repeatedly observed with JWST since the beginning of its science operation as part of multiple programs. An RGB mosaic image using the NIRCam data is shown in Figure~\ref{fig:mosaic}, where we highlight the positions of the two images. Hereafter, we refer to {the intrinsic system in the source plane} as ``\id'', short for ``Abell~2744 Metal-poor Object REdshift 6'', {and the two images in the imaging plane as \ida\ and \idb.}

We analyzed public NIRCam WFSS data from Cycle 2 program ``All the Little Things'' (GO 3516; PIs: Matthee \& Naidu)\cite{naidu24}, which targeted the Abell~2744 field using the F356W+GRISM R configuration. These observations provide wavelength coverage of $3$--$4\,\mu$m at the position of \id. Spectra at both image positions are obtained from two sets of exposures, taken at slightly different position angles (PAs; $56^\circ$ and $60^\circ$). For each source, we extracted spectra separately for the two PAs, and also extracted a combined spectrum at each image position using data from both PAs.

The spectra of the two lensed images (hereafter \ida\ and \idb) are shown in Figure~\ref{fig:spectrum}. 
Along the spectral trace at the position of \idb, we clearly detected an emission line at $\sim3.271\,\mu$m, matching the expected wavelength of \hb\ based on the \ly\ redshift measured by VLT/MUSE. We detected an emission line at the same wavelength in individual spectra at both PAs (Supplementary Materials, ``NIRCam WFSS Reduction''). The presence of the emission line at the same wavelength in both spectra firmly attributes the origin to the \hb\ line of \id, not spectral contamination from other surrounding sources. Similarly, the \hb\ line is detected in the spectrum of \ida, the counter image of \idb. While \ida\ is located in a relatively crowded region near two foreground galaxies, the spectra extracted from both PA data sets reveal \hb\ at the expected position as well. Lastly, we combined the 1-dimensional spectra of \ida\ and \idb, correcting each for its respective magnification factor to recover intrinsic flux densities, and stacked them using inverse-variance weighting (Figure~\ref{fig:spectrum}, bottom panel). 

{The comparison of the \hb\ and \ly\ lines in velocity space is shown in Figure~\ref{fig:lya_hb}. The two lines are consistently aligned, with a negligible offset of $\Delta v=$\,\dvhbly. We find that the \hb\ line exhibits broadening. This is explained by the convolution effect by the source morphology along the dispersion direction, which is inherent in slitless spectroscopy. The intrinsic \hb\ line width inferred from our forward modeling is consistent with that of \ly\ (``\hb\ Emission Line Profile'').}

\subsubsection*{Oxygen Abundance}
Intriguingly, none of the reduced spectra reveal the \oiii-doublet lines (rest-frame $\lambda\lambda4960,5008$) at the expected wavelength. The doublet lines are usually very bright in star-forming galaxies with non-zero metallicity {and high-ionization sources, especially those seen at these redshifts} ($R3\equiv \oiiir$/\,\hb\,$\gg1$). The absence of \oiii\ immediately indicates that \id\ harbors a very low-metallicity, near pristine interstellar medium. Using the conservative upper limit derived from the stack spectrum and strong line calibrations derived from photo-ionization models (``Inference of Oxygen Abundance''), we placed a firm upper limit on the oxygen abundance, {12+\logoh\,$<$\,\oxygen\ (\oxygentwosigma), or $<$\,\oxygensun\ (\oxygensuntwosigma) of solar metallicity, at \qsigma\,$\sigma$ (2\,$\sigma$).}

The inferred upper limit to Oxygen abundance is shown in Figure~\ref{fig:logoh} and compared with high-redshift galaxies observed with JWST \cite{nakajima22,laseter23,morishita24b,chakraborty24,nakajima25,maiolino25}. Of particular interest is LAP1, a previously reported metal-poor star complex at $z=6.639$ \cite{vanzella23}. A recent study \cite{nakajima25} followed up one of the clumps, LAP1-B, with JWST/NIRSpec and detected the \oiii-doublet lines along with other emission lines originated by heavy elements. Compared to LAP1-B, \id\ is $>0.3$\,dex lower in $\log R3$ and $>0.3$\,dex lower in \logoh, establishing itself as the least enriched galaxy so far discovered. For its stellar mass, \id\ is placed $\sim1$\,dex below the mass-metallicity relation derived at the corresponding redshift \cite{morishita24b}. All measurements are reported in Table~\ref{tab:lines}.

\subsubsection*{Photometric Properties}
Besides the spectroscopic features revealed above, \id\ has several exceptional characteristics, as inferred from the 20-band NIRCam photometry. Those include: (i) low stellar mass, (ii) blue UV spectral slope $\beta_{\rm UV}$, (iii) high specific star formation rate, (iv) compact morphology {, and (v) large \hb\ equivalent width.}

\id\ is very low-mass ($M_*=$\,\mstele), UV faint ($M_{\rm UV}=$\,\muv), and blue ($\beta_{\rm UV}=$\,\uvbeta), {derived from spectral energy distribution (SED) modeling}. {The very low-mass was derived assuming a Salpeter initial mass function (IMF); the mass estimate could be even smaller if a top-heavy Pop~III IMF was assumed \cite{schaerer02}.} 
The derived UV slope is bluer than the lower limit produced by standard stellar population models or nebular continuum ($\approx-2.5$). This likely requires enhanced massive star populations \cite{bouwens10,eldridge17,schaerer25} {with a non-zero ionizing-photon escape fraction; otherwise, the observed SED would be more dominated by the nebular emission, resulting in a redder slope \cite{katz24,topping24}. Its exceptionally high \ly-escape fraction, \lyesc, inferred from the measured \ly\ and \hb\ fluxes (``NIRCam WFSS Data Reduction''), supports the idea.} Using the \hb-based star formation rate (SFR), a high specific star formation rate is obtained, {SFR/$M_*$\,=\,\sSFR\,Gyr$^{-1}$.} {Achieving such a high value requires that the system be in a feedback-free, pre-supernova phase \cite{ferrara25}. This requirement seems to be} consistent with the hypothesis that \id\ is currently undergoing its first burst of star formation without having been pre-enriched. It is also interesting to note that the inferred stellar mass is compatible with what could be expected in an atomic cooling halo with modest star-formation efficiency. This would be in line with star formation in a pristine gas cloud. 

Despite the large magnification, \id\ seems to consist of a single smooth component in the image plane, without showing multiple components in the image plane. This makes \id\ distinct from previously reported strongly lensed objects that show extended arc-like features or a number of associated clumps \cite{welch22,fujimoto24}. {From our source-plane image reconstruction, we measured its intrinsic effective radius to be $R_e=$\,\relenst, resulting in moderately high stellar mass and SFR surface densities, $\Sigma_*=$\,\SIGMAS\ and $\Sigma_{\rm SFR}=$\,\SIGSFR. Although \id\ is characterized by very small physical size, the extended feature seen in the image plane suggests that \id\ is not a single globular cluster or an active galactic nucleus (AGN). The absence of significant line broadening in \ly\ is also consistent with the absence of AGN.}

Lastly, we note that our SED modeling indicates a moderately high contribution of nebular emission, which is expected {in environments hosting a young (a few $10^6$\,yr), high temperature ($\simgt5\times10^4$\,K) source \cite{panagia03,katz24}.}
Using the continuum flux density derived by our SED modeling, we find very high \hb\ rest-frame equivalent width, \ewhb\,$=$\,\ewhberr, which can be reached only by massive stars $(> 50$--$100\,M_\odot)$ of young ages ($<4$\,Myr) \cite{schaerer02,schaerer25}. {The light-weighted age inferred from the SED analysis, \tage, is consistent with this timing estimate.} All the observations here support the hypothesis that \id\ hosts primordial, massive star formation on a very small physical scale, making it the best metal-free galaxy candidate observed so far.

\subsubsection*{Implications for Cosmological Models}
\begin{figure*}[!htb]
\centering
	\includegraphics[width=0.49\textwidth]{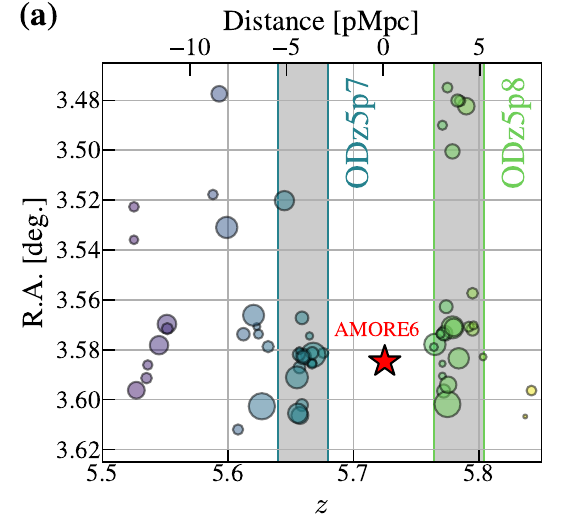}
	\includegraphics[width=0.49\textwidth]{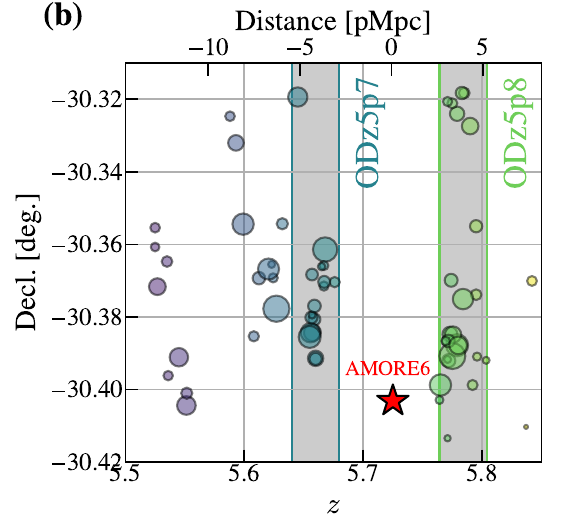}
	\includegraphics[width=0.75\textwidth]{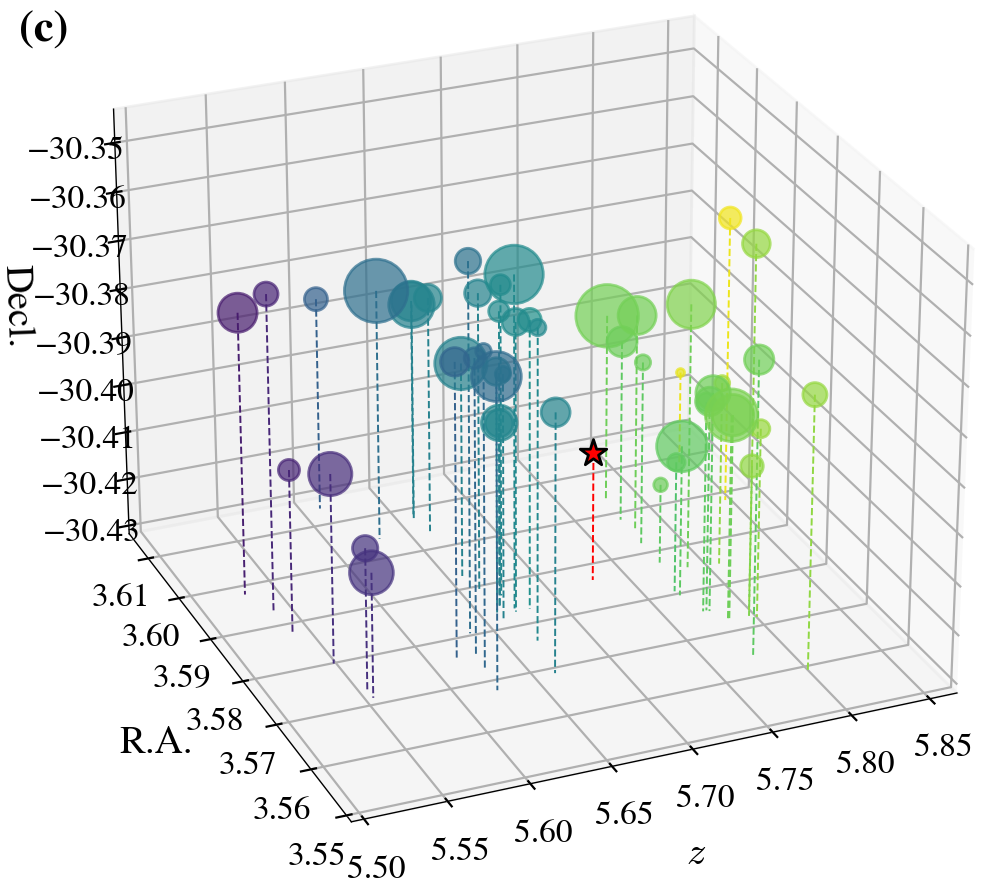}
	\caption{
    {\bf Spatial distributions of \id\ and surrounding galaxies.}
    \id\ (red symbols) is located approximately in the middle of two previously reported galaxy overdensities\cite{morishita24c} at $z=5.66$ (ODz5p7) and $z=5.78$ (ODz5p8). The previous study identified no spectroscopic sources in between the two overdensities. The symbol size of galaxies (circles) is scaled by the corresponding stellar mass. Data are adapted from \cite{morishita24c}. 
    }
\label{fig:od}
\end{figure*}

\id\ is observed about one billion years after the Big Bang. Previous studies have found that galaxies were already enriched at high redshift, within the first few hundred million years \cite{nakajima23,sanders24,laseter23,stiavelli23,Heintz2023,curti23,morishita24b}. Although the finding of a near-pristine object at a relatively late time in cosmic history could be seen as surprising, the existence of such pristine halos is not entirely unexpected {\cite{FerraraLateIII,tornatore07,TSS,BovillPopIII,visbal20,venditti23}}.
Indeed, the existence of objects like \id\ is a powerful validation of our galaxy formation and clustering models. We note that near the redshift of \id, there are two {\it evolved} overdensities of galaxies along the same sightline, at $z=5.66$ and $5.78$ \cite{morishita24c}. As shown in Figure~\ref{fig:od}, \id\ falls in the ``empty'' space, where the aforementioned study found {\it zero} spectroscopically confirmed galaxies existing between the two overdensities{, with a physical separation of $\sim$a few Mpcs}. One possible interpretation is that the gravitational potential towards both overdensities might have {delayed the formation of stars in the space between them and kept the environment therein relatively clean of galactic pollution until this relatively late cosmic time \cite{stiavelli10,xu13}.} 

The identification of a pristine object represents the confirmation of our models predicting that elements heavier than Helium are made in stars and is a basic validation of the cosmological models for the growth of structure in the Universe. The discovery of \id\ is encouraging, as it showcases the presence of near-pristine galaxies well within the reach of JWST. While JWST has shown that many of the details of our models for galaxy formation need to be revised, it has enabled us to confirm one of the basic ideas behind those models.




\begin{figure*}[!htb]
\centering
	\includegraphics[width=1\textwidth]{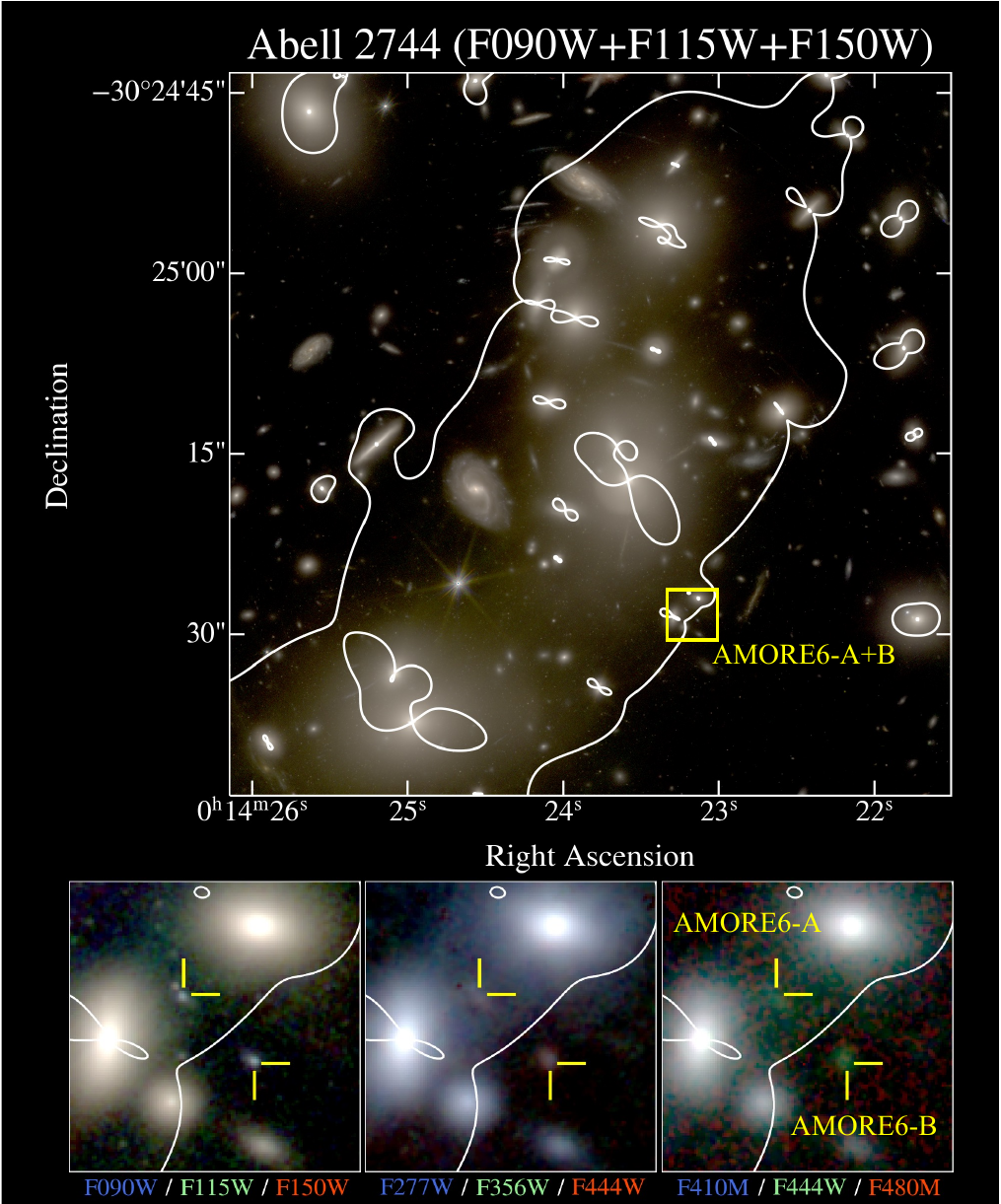}
	\caption{
    {\bf Mosaic image of the Abell~2744 field.}
    ({\bf Top:}) The observed position of the \id-A+B system is indicated in the mosaic image (square). The double image system straddles the critical curve predicted from the latest gravitational lens model\cite{bergamini23b} (white lines). North is up.
    ({\bf Bottom:}) Zoomed-in cutout images of the \id-A+B system in the size of $4''\times4''$, in pseudo RGB colors combined with various NIRCam filters. 
    }
\label{fig:mosaic}
\end{figure*}

\begin{figure*}[!htb]
\centering
	\includegraphics[width=0.999\textwidth]{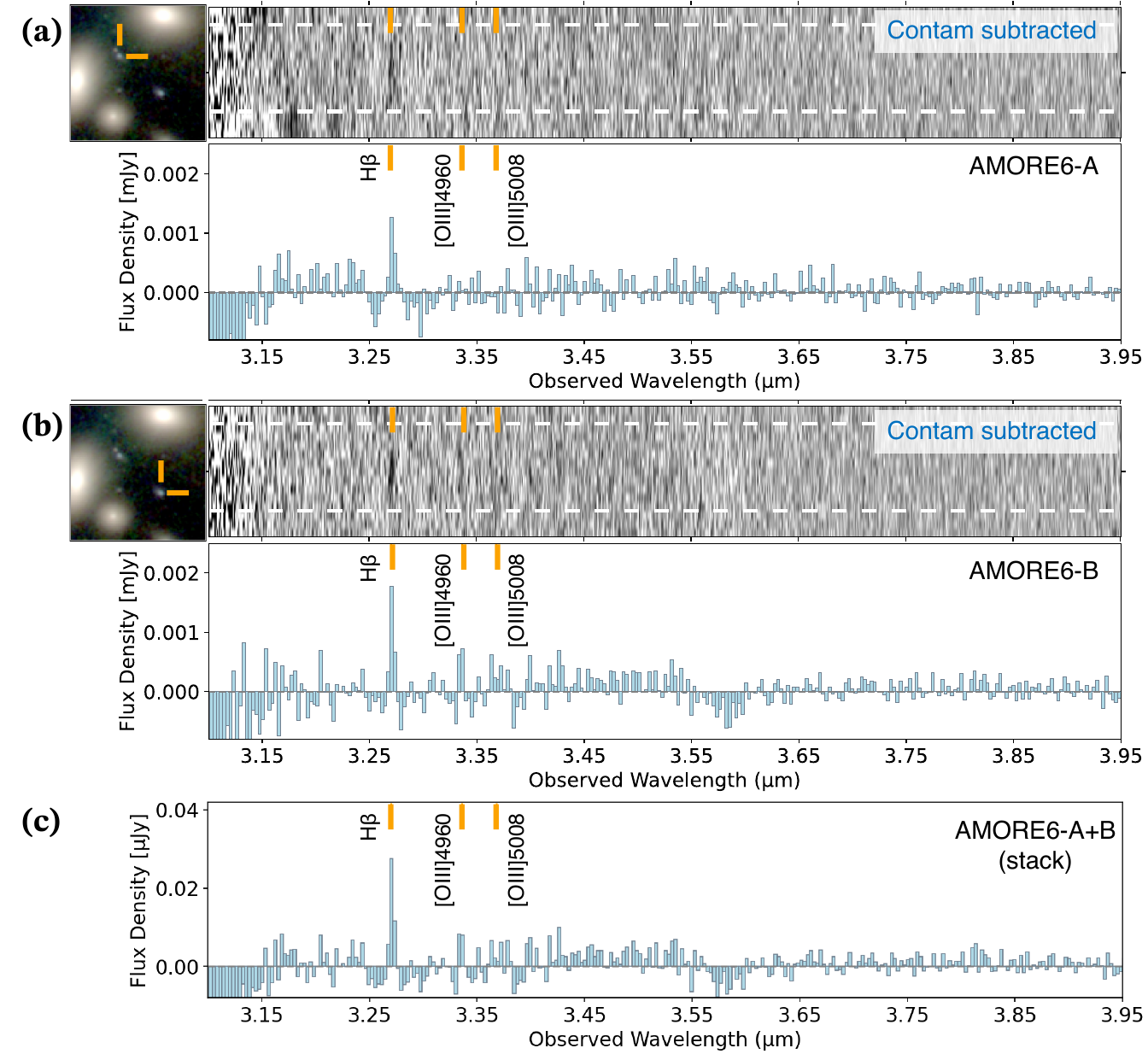}
	\caption{
    {\bf WFSS spectra of \id.}
    {\bf (a)} Observed WFSS spectrum extracted at the position of \ida\ (cutout image in the size of $3.\!''2\times3.\!''2$). The top row shows the extracted 2-dimensional spectrum, with contamination being subtracted (see ``NIRCam Wide Field Slitless Spectroscopy Data Reduction''). {Horizontal dashed lines indicate the positions $\pm0.\!''25$ from the source extraction trace center (see also Figure~\ref{fig:spectrum_ind}).} The bottom row shows the extracted 1-dimensional spectrum, {binned at the spectral {bin} size of 30\,\AA, which corresponds to approximately three native spectral elements
    }. The wavelengths of the \hb\ and \oiii-doublet lines are indicated by vertical bars. 
    {\bf (b)} Same as the top panel but for \idb.
    {\bf (c)} 1-dimensional spectrum of the \id-A+B stack spectrum. Each spectrum is normalized by the magnification factor, and thus the final spectrum represents the ``unlensed'' flux.
    }
\label{fig:spectrum}
\end{figure*}

\begin{figure*}[!htb]
\centering
	\includegraphics[width=0.99\textwidth]{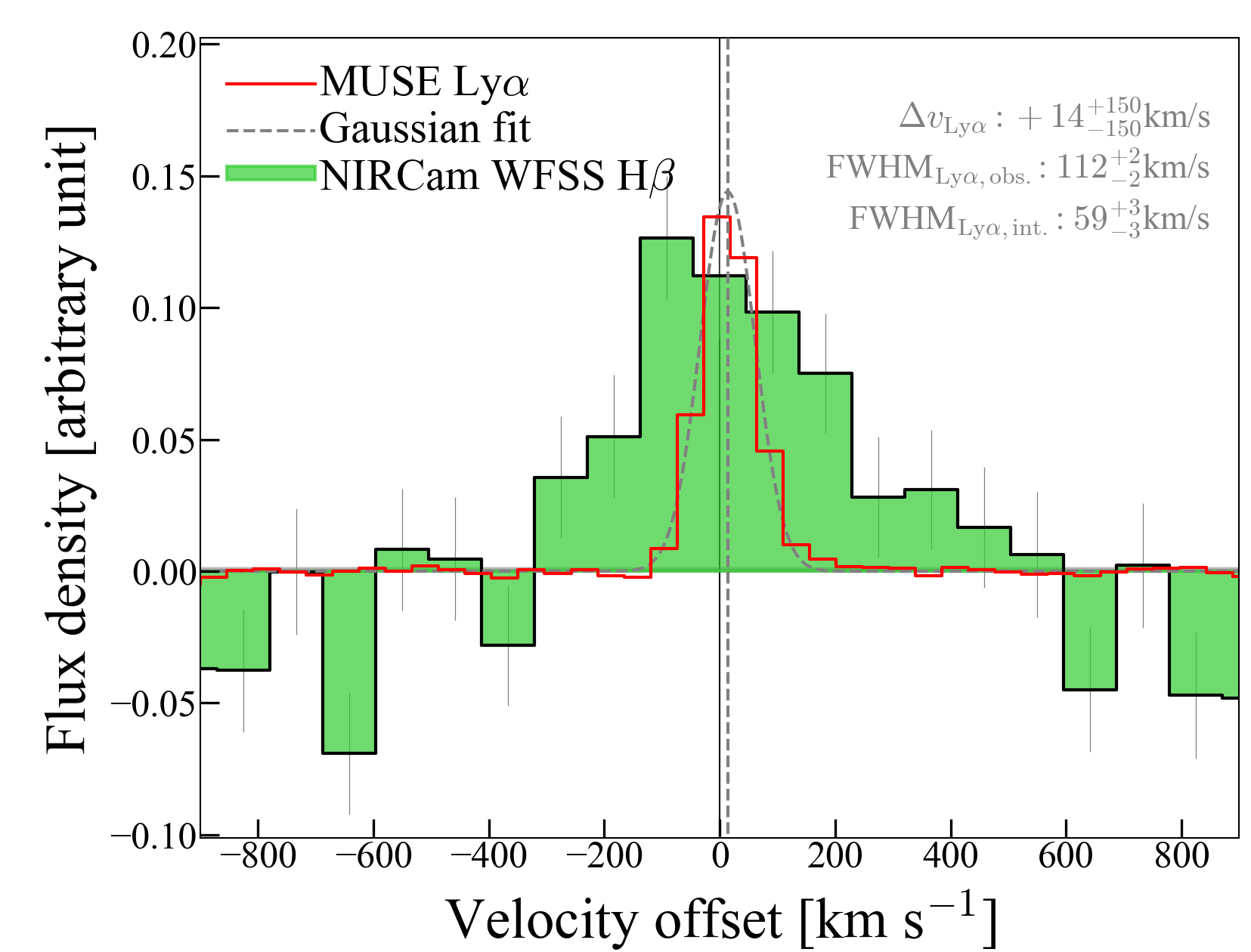}
	\caption{
    {\bf Line velocity offset between \ly\ and \hb.}
    The observed \ly\ from VLT/MUSE (red solid line) and \hb\ from JWST/NIRCam WFSS (green filled, with 1\,$\sigma$ error bars, {in the native spectral dispersion 10\,\AA\,/\,pixel}), normalized to an arbitrary unit, are shown.
    {The \hb\ line broadening is primarily due to line-spreading from slitless spectroscopy, such that convolution effect of the source morphology along the dispersion direction.
    The \ly\ line is narrow but shows asymmetry, with a slightly enhanced flux on the red side of the peak wavelength compared to the fitted single gaussian profile (gray dashed line), which is characteristic to the line at this redshift.}
    }
\label{fig:lya_hb}
\end{figure*}

\begin{figure*}[!htb]
\centering
	\includegraphics[width=0.999\textwidth]{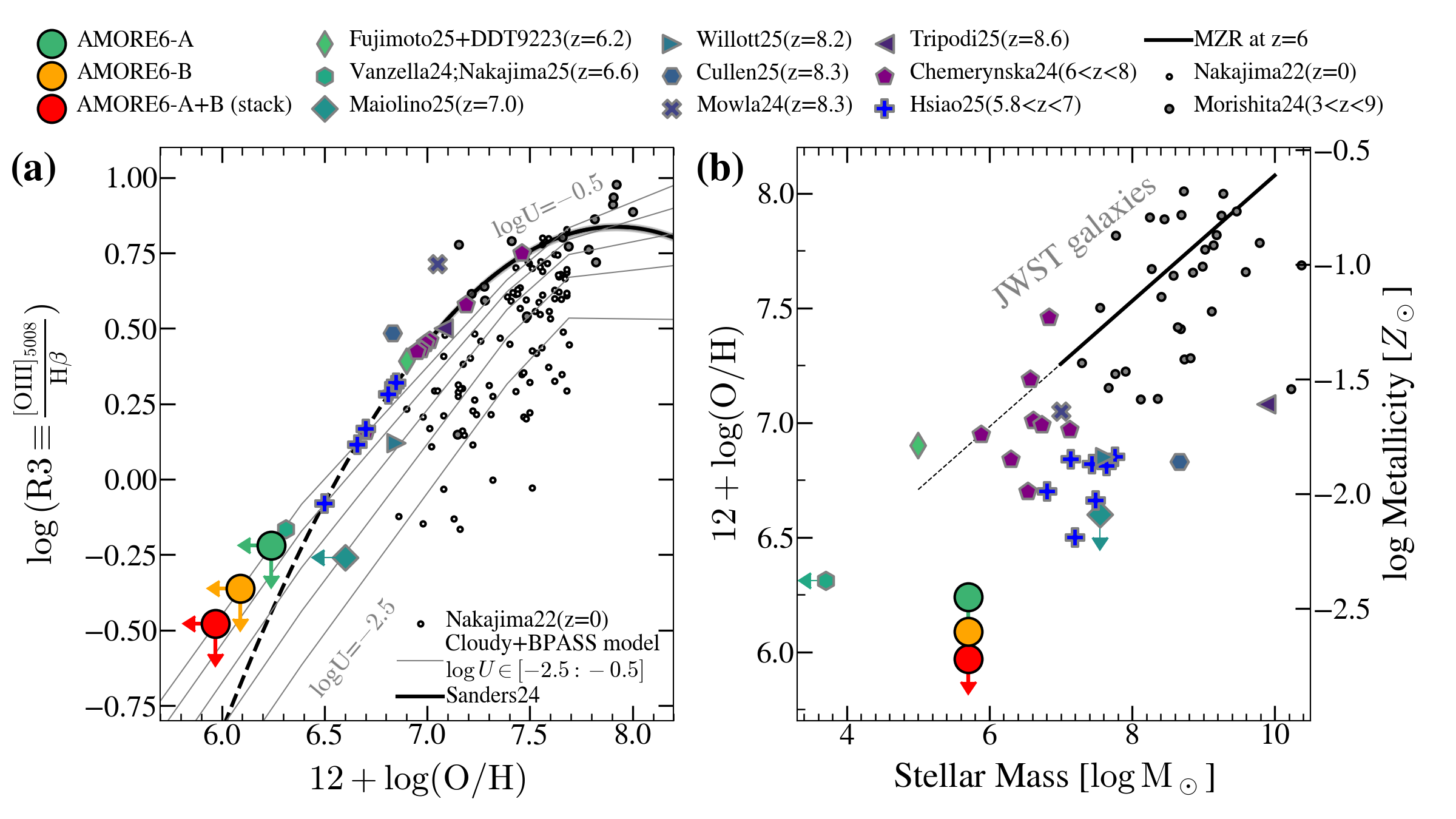}
	\caption{
    {\bf Oxygen abundance of \id.}
    {\bf (a)} $R3\equiv\oiiir/$\hb\ measurements of \ida, \idb, and \id-A+B stack ({$3\,\sigma$ upper limits}). Photo-ionization models (gray solid lines in the back ground) are used to infer \logoh\ as a function of observed $R3$. For comparison, galaxies at $3<z<10$ from studies in the literature are shown {\cite{chemerynska24,tripodi24,nakajima25,maiolino25,mowla24,hsiao25,fujimoto25,willott25,cullen25}}, using the reported $R3$ and \logoh\ measurements (small symbols). {Metallicity measurements of $z\sim0$ extremely metal-poor galaxies \cite{nakajima22} and $z=3$--10 galaxies \cite{morishita24b}, both using the auroral \oiiif\ line (direct method), are also shown.}
    {\bf (b)} \id\ on the stellar-mass metallicity plane. The same literature galaxies at $z>3$ are shown, along with the mass-metallicity relation at $z=6$ \cite{morishita24b}.
    }
\label{fig:logoh}
\end{figure*}

\clearpage 

%
\bibliography{science_template} 
\bibliographystyle{sciencemag}

%
%
%
%
%
%


\section*{Acknowledgments}
We are grateful to the reviewers for their careful reading and constructive comments, which improved the manuscript. We acknowledge the teams of the JWST observation programs, IDs~1324, 2561, 2756, 2883, 3516, 3538, 3990, and 4111 for their dedicated work in designing and planning these programs and for generously making their data publicly available. We would like to thank Piero Rosati and Claudio Grillo for the discussion about the Abell~2744 lens modeling and their efforts made for the model construction, Eros Vanzella for his useful comments on the paper, Andrea Ferrara, Stefano Carniani, and Keunho Kim for constructive discussions, and Lilan Yang for the help with lenstruction. 

\paragraph*{Funding:}
T.~M. received support from NASA through the STScI grants HST-GO-17231 and JWST-GO-3990. Z.~L. is supported by the Japan Society for the Promotion of Science (JSPS) through KAKENHI Grant No. 24KJ0394. M.~S. acknowledges support for this work under NASA grant 80NSSC22K1294. P.B. acknowledges financial support through grant PRIN-MIUR 2020SKSTHZ and support from the Italian Space Agency (ASI) through contract ``Euclid - Phase E'', INAF Grants ``The Big-Data era of cluster lensing'' and ``Probing Dark Matter and Galaxy Formation in Galaxy Clusters through Strong Gravitational Lensing.''

\paragraph*{Author contributions:}
T.M. led the JWST reduction and the main data analysis and wrote the manuscript. 
Z.L. led the JWST NIRCam WFSS reduction and the analysis and wrote the manuscript. 
M.S. contributed to the analysis and to the scientific discussion and wrote the manuscript.
T.T contributed to the scientific discussion and wrote the manuscript.
P.B. contributed to the lens analysis of the main target and calculated magnification factors at the positions of the two images. 
Y.Z. contributed to the photoionization models used for the oxygen abundance inference. 
All authors contributed to the discussion of results and to the final editing of the manuscript.

\paragraph*{Data and materials availability:}
The JWST and HST data used in the manuscript are publicly available through MAST (\url{https://archive.stsci.edu}). The specific JWST data are accessible at \cite{https://doi.org/10.17909/kr0e-8j85} and \cite{https://doi.org/10.17909/sezc-eb95}. The HST imaging data are available at \url{https://cosmos.phy.tufts.edu/~danilo/HFF/Home.html}. The VLT/MUSE data are available at the ESO Science Archive Facility (\url{https://archive.eso.org/scienceportal/home}), or via DOIs \cite{https://doi.org/10.18727/archive/41,https://doi.org/10.18727/archive/42}.

\input{materials}


\newpage


\renewcommand{\thefigure}{S\arabic{figure}}
\renewcommand{\thetable}{S\arabic{table}}
\renewcommand{\theequation}{S\arabic{equation}}
\renewcommand{\thepage}{S\arabic{page}}
\setcounter{figure}{0}
\setcounter{table}{0}
\setcounter{equation}{0}
\setcounter{page}{1} 


\begin{center}
\section*{Supplementary Materials for\\ \scititle}

\input{authors}
\end{center}

\input{materials}
\newpage


\subsection*{Materials and Methods}
Throughout the manuscript, we adopt a standard $\Lambda$CDM cosmology with $\Omega_{\Lambda}=0.7$, $\Omega_{m}=0.3$, and $H_0=70$\,km\,s$^{-1}$\,Mpc$^{-1}$. All magnitudes are given in the AB magnitude system \cite{oke83,fukugita96}. The solar metallicity value 12+\logoh$=8.69$ \cite{asplund09} is adopted throughout the paper. {Unless specifically noted, $1\,\sigma$ error is quoted for measurements and $3\,\sigma$ upper limit is quoted for non-detections throughout the paper.}

\subsection*{NIRCam WFSS Data Reduction}\label{sec:wfss}
We analyzed public NIRCam Wide Field Slitless Spectroscopy (WFSS) data from the Cycle 2 program ``All the Little Things'' (GO 3516; PIs: Matthee \& Naidu) \cite{naidu24}, which provides F356W coverage at the position of \id\ over the wavelength range of $3$--$4\,\mu$m. We adopt the median filter technique for continuum subtraction, which has been widely adopted in the literature \cite{Kashino23_grism, Sun23_grism, liu24}. This approach models the observed continuum using a moving window (or ``kernel''), and subtracts it from the pre-processed 2D spectrum (that is, flat-fielded and background-subtracted) to effectively isolate the emission line signals.

The spectra were obtained at two slightly different PAs, approximately $56^\circ$ and $60^\circ$, {all with Grism R, Module A}. To control the effects of PA-dependent spectral overlap from nearby sources (``contamination''), we extracted spectra at the image positions of \ida\ and \idb\ from each PA individually, as well as from the combined data set including both PAs. This approach allowed us to assess the impact of contamination and to ensure that the detected emission lines are not driven by spurious features present in only one of the orientations. Each PA data set has a total on-source exposure time of 24,738\,sec for both positions of \ida\ and \idb.

{For \ida\ and \idb, there are three foreground galaxies that contribute to spectral overlap. Two of those contaminants, C1 and C2, are near \ida, both H-band apparent magnitude $m_{\rm H}\sim21$\,mag, as seen in the cutout image (Figure~\ref{fig:spectrum_ind}). Another one, C3, is a bright galaxy ($m_{\rm H}\sim20$\,mag), located $0.\!'4$ away from \id\ to the South-East, the primary contaminant in the \idb\ PA=56$^\circ$ spectrum.
}

The reduced spectra of \id\ are shown in Figure~\ref{fig:spectrum}, where we present the combined spectrum of each source. At the position of \idb, we clearly see the detection of an emission line at $\sim3.271\,\mu$m, corresponding to \hb\ at $z\sim\zspec$, consistent with the redshift measured from \ly\ by VLT/MUSE. 
{The \ly\ line is narrow but shows slight asymmetry, with a reduced flux on the blue side of the peak wavelength. This is a characteristic feature of the line at this redshift, due to non-zero neutral hydrogen fraction, thus being a unique determining factor of its redshift. The photometric redshift, derived with the NIRCam photometric data (see below), is consistent with the redshift solution.}

In Figures~\ref{fig:spectrum_ind} and \ref{fig:spectrum_ind_idb}, we show the individual spectra of \ida\ and \idb\ extracted at each PA. The \hb\ line is detected at the position of \ida, in both PAs. Although \ida\ is located closer to foreground sources and mildly suffers from spectral contamination, \hb\ is detected in both spectra. The spectrum of \idb\ taken at PA\,$=56^\circ$ suffers from significant contamination by a foreground cluster galaxy, whereas the spectrum at PA\,$=60^\circ$ is mostly contamination-free (Figure~\ref{fig:spectrum_ind_idb}). The presence of the emission line at the same wavelength in all spectra firmly attributes the origin to the \hb\ line of \id.

To increase the signal-to-noise ratio (SNR), we combined the 1-dimensional spectra of \ida\ and \idb. Each spectrum was normalized by the magnification factor estimated at the position. The normalized spectra are then stacked by weighting with the inverse variance. The combined spectrum is shown in the bottom panel of Figure~\ref{fig:spectrum}. The spectral region around \hb\ of the combined spectrum is also shown in Figure~\ref{fig:lya_hb}, compared to \ly\ extracted from the existing VLT/MUSE data. The \ly\ flux was extracted at the positions of \ida and \idb\ and then integrated. The broader line width observed in the WFSS spectrum than in MUSE \ly\ is attributed to two factors: the spectral resolutions ($R\sim1500$ and $\sim3000$) and the convolution effect by the source morphology, which is unique to slitless spectroscopy. {The flux ratio \ly-to-\hb\ for \ida+B is \lytohb, which is high but still smaller than $23.5$ for Case B ($n_e=10^2$\,cm$^{-3}$, $2\times10^4$\,K), meaning that \lyesc\ of \ly\ photons escape. The cause of the reduced escape fraction, especially in a supposedly dust-free environment, remains unconcluded with the available data; a possible explanation would be resonant scattering by neutral gas, where some scattered \ly\ photons to a large extent fall below the surface brightness detection limit.}

The line profile of each emission line of interest in the extracted 1d spectrum was modeled with a Gaussian function. The amplitude and line width parameters are included for each line model, in addition to one global redshift parameter. Three lines, \hb, $\oiiib$, and $\oiiir$, were attempted to fit simultaneously. The fit is iteratively repeated for 5000 times through the Markov chain Monte Carlo method using {\sc emcee} \cite{foreman14}. The integrated flux of each line is estimated by integrating the corresponding Gaussian model. 

Flux errors are calculated in three different ways: one is by taking the differences between the 16th and 84th percentiles of the MCMC model. The second approach is to sum the observed flux density error weighted by the amplitude of the Gaussian model in quadrature. The third approach is to integrate the observed flux density error over a fixed line width (here set to 150\,km/s) using a uniform weight. The largest flux error among those calculated with the three approaches is reported and adopted in the main analysis. 

The fitting results are shown in Figure~\ref{fig:lines}. We detect \hb\ in the spectra of \ida, \idb, and the \ida+B stack at SNR of \snra, \snrb, and \snr, respectively. Neither one of the \oiii-doublet lines is detected at SNR\,$>3$ in any spectra. The measured line fluxes and flux upper limits are reported in the main article Table~\ref{tab:lines}. {The derived redshift from the stacked spectrum, $z=5.7253\pm0.0005$, is consistent with those presented in \cite{mahler18} ($z=5.7255$) and \cite{bergamini23} ($z=5.7256$) within $1\,\sigma$ uncertainties.}

We find that the spectrum of \idb\ shows six consecutive non-zero spectral pixels near the wavelength of $\oiiib$ {(Figure~\ref{fig:lines})}. This corresponds to two consecutive spectral pixels in the resampled spectrum (presented in the main text Figure~\ref{fig:spectrum}). {Through a careful inspection of the feature in the two separate roll spectra, we found that this feature is only visible in the spectrum at PA\,$=56^\circ$. This specific spectrum is significantly contaminated by a bright foreground galaxy (marked as C3 in Figure~\ref{fig:spectrum_ind_idb}), resulting in several similar features in the cleaned spectrum across the wavelength range.} In addition, the same feature is not consistently seen in the spectrum of \ida, {which would still have been detectable for the noise level if the feature originates from the source}. Lastly, the $\oiiib$ line is about 3 times fainter than its doublet counterpart, $\oiiir$ \cite{storey00}. The absence of $\oiiir$ in all the spectra analyzed here supports the interpretation that the observed positive pixels at the wavelength of $\oiiib$ are an artifact, {likely a residual of spectral contaminants}.

\subsection*{{\hb\ Emission Line Profile}}\label{sec:hb}
{The detected \hb\ line exhibits significant broadening, compared to that of \ly. This is explained by the convolution effect of the source morphology along the dispersion direction, which is inherent in slitless spectroscopy:
\begin{equation}
\mathrm{FWHM}_{\mathrm{obs}} = 
\sqrt{\mathrm{FWHM}_{\mathrm{intrinsic}}^2 
      + \mathrm{FWHM}_{\mathrm{lsf}}^2},
\end{equation}
where $\mathrm{FWHM}_{\mathrm{lsf}}$ is the combined line-broadening effect originating in the instrumental line-spread function (LSF) and the source morphology along the dispersion direction. The source appears spatially extended for all dispersion position angles and image positions, suggesting that the observed line broadening is primarily due to morphological rather than intrinsic kinematic effects.

To quantify the morphology effect along the dispersion direction, we used the NIRCam F356W imaging, rotated it to match the PAs of the Grism R observations in Module A, such that the x-axis aligns with the dispersion direction. We then collapsed the rotated image along the cross-dispersion direction and measured the FWHM of the resulting one-dimensional profile. 
For AMORE6-A, the measured FWHM of the collapsed profile is $\sim0.\!''22$ at both angles, or $33$\,\AA\ when converted to the wavelength unit using the native pixel scale. For AMORE6-B, the measured FWHM is similar, with $\sim0.\!''22$ for both angles, or $33$\,\AA. 
We note that the light profile measured this way includes the PSF effect of the NIRCam F356W imaging, not WFSS. However, the empirical PSF size for the imaging filter is $0.\!''12$ \cite{jdox}, which is smaller than the broadening effect of the NIRCam WFSS at the wavelength of interest, $\sim0.\!''15$ \cite{Greene17}. Taking this into account, it is suggested that the observed line profiles are dominated by the morphology dispersion effect. 

To quantify the intrinsic line width, we attempt to forward-model the observed line profile. We follow the same approach in our gaussian line fitting presented above, but with the gaussian model convolved with the F356W imaging kernel at each iteration. From this forward modeling, we find $\mathrm{FWHM}_{\mathrm{intrinsic}}=5.4_{-3.2}^{+4.2}$\,\AA, or $49_{-29}^{+38}$\,km/s, for AMORE6-A and $7.7_{-3.9}^{+3.4}$\,\AA, or $70_{-35}^{+31}$\,km/s, for AMORE-B (Figure~\ref{fig:lsf}).
Given the similar LSF profile for both PAs, to increase signal-to-noise ratios, we use the stacked spectrum for each image. The inferred intrinsic velocity broadening is consistent with the \ly\ line width ($\mathrm{FWHM}_{\mathrm{Ly\alpha, intrinsic}}=69.3_{-0.7}^{+0.8}$\,km/s after subtracting the instrumental broadening component) within the $1\,\sigma$ uncertainties.
}

\subsection*{Imaging Data Reduction}\label{sec:image-jwst}
\id\ is located behind the galaxy cluster Abell~2744 at $z=0.308$. This field has been repeatedly observed with JWST since the beginning of its science operation in multiple cycles (see below). We utilize JWST/NIRCam imaging data available in this field from multiple programs: GLASS-JWST (ERS 1324; PI: Treu) \cite{Treu22}, UNCOVER (GO 2561; PI: Labbe) \cite{bezanson22}, DD 2756 (PI: Chen) \cite{chen22}, GO 2883 (PI: Sun), All the Little Things (GO 3516; PIs: Matthee \& Naidu) \cite{naidu24}, GO 3538 (PI: Iani), BEACON (GO 3990; PI: Morishita) \cite{morishita24beacon}, and Medium Bands, Mega Science (GO 4111; PI: Suess) \cite{suess24}.

All NIRCam images of 20\,bands are reduced in a consistent way, following\cite{morishita24beacon}. The JWST pipeline version used for the analysis is ver1.18.0 and the pmap context 1298. We process the images to have the same pixel grid, with the pixel scale of $0.\!''0315$, after registering to the astrometry defined by GAIA DR3. We supplement the data with publicly available deep HST/ACS F435W, F606W, and F814W images \cite{shipley18}. 

In summary, we utilize 20 NIRCam and 3 HST bands. The $5\,\sigma$ point source limiting magnitudes: HST-F435W (28.7), F606W (28.1), F814W (27.7), NIRCam-F070W (28.5), F090W (28.9), F115W (28.7), F140M (27.9), F150W (28.8), F162M (27.9), F182M (28.1), F200W (28.9), F210M (28.0), F250M (28.0), F277W (29.1), F300M (28.1), F335M (28.1), F356W (29.1), F360M (28.2), F410M (28.6), F430M (27.7), F444W (28.9), F460M (26.8), F480M (27.2).

\subsection*{\bf Strong Lens Model}\label{sec:lens}
{
The presented analysis relies on the high-precision strong-lensing model described in \cite{bergamini23b}, currently among the most advanced lens models available for the galaxy cluster Abell 2744. This model exploits JWST and VLT/MUSE spectro-photometric observations to identify a large sample of secure multiple images, as well as a pure and complete catalog of cluster member galaxies. Specifically, it is constrained by point-like positions of 149 multiple images spanning a redshift range between 1.03 and 9.76, including the \id\ system. The model achieves a remarkable precision of $0.\!''43$ in reproducing the positions of the multiple images, which improves to $0.\!''17$ for the AMORE6 system. 

From this model, we estimated magnification values of $\mu_{\rm A}=$\,\magnifaerr\ and $\mu_{\rm B}=$\,\magnifberr\ for \ida\ and \idb. These values and their associated statistical uncertainties are derived from 5000 realizations of the lens model, generated by randomly extracting parameter samples from the model MCMC chains and computing, for each realization, the magnification at the model-predicted positions of the multiple images. The reported values correspond to the 16th, 50th, and 84th percentiles of the resulting magnification distributions. For comparison, we also estimated the magnification values from the best-fit lens model---corresponding to the set of parameters that minimizes the difference between the observed and model-predicted positions of the multiple images---obtaining values of 44.6 and 86.5 for \ida\ and \idb, respectively.

We note that, although the adopted model represents the current state of the art for Abell 2744, the magnification estimates may still be affected by non-negligible systematic uncertainties that are not considered in our analysis. In fact, from the WFSS spectral analysis we find the \hb\ line ratio of $1.44 \pm 0.42$ between \idb\ and \ida, whereas the ratio of magnification factors is $1.98 \pm 0.29$. Similarly, photometric analysis in the F115W image measures the ratio to be $1.38 \pm 0.05$, although the flux measurement of \ida\ suffers from foreground contamination. Conservatively, we could derive the magnification factor of \idb\ using $\mu_{\rm A}$ multiplied by a factor 1.4 i.e. $\mu_{\rm B}\sim$\,\magnifbfroma, assuming that the magnification measurement for \ida\ is relatively more robust being farther from the critical line. The influence of magnification uncertainties is limited to absolute measurements, such as stellar mass, star formation rate, UV absolute magnitude, and effective radius; other measurements such as line flux ratio, oxygen abundance, spectral slope, specific star formation rate, age, and stellar mass and star formation surface densities are not affected.
}

\subsection*{Photometry}\label{sec:data-jwst}
The reduced images were used for photometry. We identified sources in the rest-frame UV detection image (F150W+F200W stack) using \textsc{Source-Extractor} \cite{bertin96} and measured flux on each image (matched to the F444W point-spread function, PSF) within a fixed aperture of radius $r=0.\!''16$\,. This study utilizes the F150W+F200W stack image for source detection, as opposed to, for example, longer wavelength filters often used in the literature. The choice was made because the former offers a $\sim2\times$ higher spatial resolution, which increases the detection completeness of faint sources in relatively crowded regions, as is the case for \id. Especially, \ida\ is located near two passive, bright, foreground galaxies at $z\sim0.5$. Besides, we detect a compact object to the North-East of \ida, which is almost blended in the longer wavelength bands. This compact object shows slightly different colors than \ida\ in the RGB images (Figure~\ref{fig:mosaic}), with photometric redshift $z_{\rm phot}=5.5_{-0.4}^{+0.3}$ obtained with {\tt EAZY} \cite{brammer08}. We did not detect emission lines along its spectral trace (``NIRCam WFSS Reduction''), while the association of the compact object remains to be confirmed in future observations. \idb\ is located in a relatively isolated region.

\subsection*{Inference of Oxygen Abundance}\label{sec:logoh}
Using the measured flux of \hb\ and upper limit on $\oiiir$, we obtain the oxygen abundance, 12+$\log$(O/H), via strong line calibration methods. The $R3$ calibrator offers a simple conversion between the line ratio $R3\equiv\oiiir/$\,\hb\ and Oxygen abundance. 

We ran photo-ionization models using {\sc Cloudy} (ver.23.01) \cite{osterbrock89,chatzikos23} and calculated $R3$ and \logoh\ in various conditions, following the methods in the literature \cite{nakajima25}.
We calculate the model for different O/H abundance ratios and ionization parameter ($\log U\in[-2.5:-0.5]$ in step of 0.5). For hydrogen density, we adopt $n_{\rm H}=1000$\,cm$^{-3}$; however, changing density to 100\,cm$^{-3}$ or 10000\,cm$^{-3}$ would not significantly affect the final result. For the ionizing source, we use {\sc BPASS} \cite{eldridge17,stanway18} binary stellar radiation assuming an instantaneous star formation history with the stellar age of 1\,Myr, upper mass cut of 100\,$M_\odot$, and the Kroupa IMF, of the same metallicity as the gas component. The model grids obtained are comparable to those presented by \cite{nakajima25} in the metallicity range of interest. 

To determine \logoh, we adopted the model of $\log U=-0.5$, the same value as was used for LAP1-B \cite{nakajima25}. This ionization parameter was chosen to make a fair comparison with LAP1-B. With this, we obtain {12+\logoh\,$<$\,\oxygen\ (\oxygentwosigma), or $<$\,\oxygensun\ (\oxygensuntwosigma) of solar metallicity, at \qsigma\,$\sigma$ (2\,$\sigma$), for the observed R3 measured from the stacked spectrum.}
Adopting $\log U=-2.5$ instead, which is the lowest acceptable per observational data in the literature, would increase the oxygen abundance but only by $\sim0.4$\,dex.  It is noted that we excluded the solution at 12+\logoh\,$\simgt8$. Allowing the range above would give us an alternative solution, 12+\logoh$>8.7$ ($>1 Z_\odot$). This is almost 2\,dex above the mass metallicity relation for the stellar mass of \id, and thus we consider it very unlikely. In addition, any stellar population model could not reproduce the observed SED (see ``Spectral Energy Distribution Modeling'') at this high metallicity.

By adopting the strong line calibration by \cite{sanders24} and extrapolating it down to the $R3$ of \id, we obtained $12+$\,\logoh\,$<\oxygensand$, or $<$\,\oxygensunsand\ of solar metallicity (\qsigma\,$\sigma$). We note that Maiolino et al. \cite{maiolino25} adopted the strong line calibration by \cite{cataldi25} to obtain their \logoh\ measurement. This explains their larger measurement than that expected from our approach (12+\logoh\,$\approx6.2$) for its relatively low $R3$ (Figure~\ref{fig:logoh}, in the main text).

{While the R3-to-log(O/H) calibration we adopt is mostly insensitive to the density at $n_{\rm H}<10^5$\,cm$^3$, the \oiii\ emission starts to decrease above the critical density of the \oiii\ upper level ($n_{\rm crit} \sim 7\times10^5$\,cm$^{-3}$). For example, by adopting $n_{\rm H}=10^6$\,cm$^3$, we would expect a $\sim0.6$\,dex increase in the \logoh\ measurement. However, such a high density is only observed near an active accreting disk \cite{maiolino25} or unresolved, extremely high density ($>10^3\,M_\odot\,{\rm pc^{-2}}$) sources \cite{topping25}. From the observed smooth extended light profile and stellar mass surface density, the high density scenario does not seem to apply to \id. Further constraints on its electron density would be required to completely discriminate the high-density scenario.}

A recent work \cite{fujimoto25} selected a candidate Pop-III galaxy using a photometric color-color diagram, further followed by a photometric redshift analysis. The candidate, GLIMPSE-16043, was recently followed up with the JWST/NIRSpec MSA (DDT9223, PI: Fujimoto). We retrieved the reduced G395M/F290LP spectrum from MAST (JWST calibration pipeline ver1.18.0 under the pmap context 1364) and applied our line-fitting analysis. The \hb\ and \oiii\ are clearly detected, yielding $\log R3=0.39 \pm 0.10$. Using the same Oxygen abundance inference above, we found 12+\logoh\,$=6.9$. The object is shown in Figure~\ref{fig:logoh} in the main text.

Shown in Figure~\ref{fig:logoh} is the mass--metallicity relation, to highlight the location of \id\ with respect to galaxies at $3<z<10$ \cite{chemerynska24,tripodi24,nakajima25,maiolino25,morishita24b,hsiao25,fujimoto25}. Compared to the mass--metallicity relation at the corresponding redshift \cite{morishita24b}, \id\ falls $>1$\,dex below for its stellar mass.

\subsection*{{Two-dimensional Light-profile Modeling}}\label{sec:size}
{We performed a source-reconstruction of \id\ in the source plane using the {\tt lenstruction} software \cite{yang20}. Using the F115W image, which corresponds to the rest-frame $1700$\,\AA, the parameterization of \id\ in the source-plane was carried out through a forward-modeling approach, by adopting an elliptical S\'ersic profile. The fitting process was carried out iteratively, where each proposed source-plane model is converted into the image-plane using the deflection maps of the best-fit lens model \cite{bergamini23b} and then convolved to the F115W PSF to mimic lensing effects, to simultaneously compare with the surface brightness of both images. The compact source to the North-East of \ida, likely a foreground source, was masked out during fitting.
In a forward-modeling approach, the provided PSF image plays a critical role. As such, we adopted 
the method presented in \cite{morishita24}. 
Briefly, we generated a set of artificial PSF images with the {\tt stpsf} tool (ver.2.0.0) \cite{stpsf}, with each having a slightly different profile controlled by the {\tt jitter} parameter. We then fitted bright stars in the F115W image with those artificial PSF images and evaluated the accuracy by measuring the total residual fluxes. For the F115W image we use in this study, we found that the one with $\sigma_{\rm jitter}=0.\!''021$ best represents the actual stellar light profiles, similar to \cite{morishita24}.

The results of the source reconstruction are shown in Figure~\ref{fig:lenst}. \id\ is characterized with the effective radius (along the major axis) of $r_e=$\,\relenst\ in the source plane, where the uncertainty also includes that in the magnification factor of the adopted lens model. We quote this size estimate as the characteristic size of \id\ throughout the main text and report in Table~\ref{tab:phys}. 

To assess potential systematics originating in uncertainties in the lens model near the image position, we repeat the reconstruction for each of the double images separately. The results of the source-plane light profile are consistent between the two images, with $r_e=$\,\realenst\ and \reblenst\ for \ida\ and \idb, respectively. 

In addition, we modeled the two-dimensional light profile in the image plane with a single S\'ersic model using {\sc galfit} software \cite{peng02}. \ida\ is characterized with $r_e=$\,\reamas\ (along the major axis) in the image plane, or \rea\ after dividing it by $\sqrt{\mu}$. \idb\ is characterized with $r_e=$\,\rebmas\ in the image plane, or $r_e=$\,\reb\ in the source plane. We note that the application of a magnification factor in this way is justified under the assumption of isotropic magnification ($\mu_{\rm tangential} = \mu_{\rm radial}$). Regardless, the mean size of the two measurements above is \re, which is consistent with the one measured with {\tt lenstruction}. 
Using $\mu=$\,\magnifbfroma\ for \idb\ instead (``Strong lensing models'') results in $r_e=$\,\rebfroma, a slightly better agreement with the size measurement for \ida. 

In Figure~\ref{fig:sizemass}, \id\ is compared with various types of object in the size-mass plane. With the measured size and stellar mass (``Spectral Energy Distribution Modeling''), \id\ is located at the upper ridge of the region dominated by local ultra-compact dwarfs and globular clusters. \id\ has the size and mass similar to individual star-forming clumps found in recent JWST observations. However, as we have seen in our source-plane image reconstruction, \id\ is resolved and composed of a single smooth component, uniquely distinguishing itself from those populations. 
}

\subsection*{SED Modeling}\label{sec:sed}
We modeled the spectral energy distribution (SED) of \id\ using the JWST+HST photometric data. We used the SED fitting code {\sc gsf} (ver.1.93) \cite{morishita19}, which allows flexible determinations of the SED by adopting binned star formation histories (often referred to as ``non-parametric''), with the age pixels set to [1, 3, 10, 30, 100, 300, 1000]\,Myrs. The fitting templates were generated using the BPASS template library (ver.2.2.1) of a single initial mass function (``imf135\_300''), with the upper cutoff mass of 300\,$M_\odot$, including binary populations. Given our interest, we only included the templates of the lowest metallicity available ($10^{-5}\,Z_\odot$). In addition, nebular templates for various ionization parameters ($\log U\in[-3.5:-0.5]$), distributed by the BPASS team, are included to account for the nebular emission and continuum component. The nebular templates are also set to the same metallicity as for the stellar templates. For dust attenuation, we adopted the SMC dust curve \cite{gordon03}. In summary, our SED modeling has 7 (amplitudes for stellar templates) + 2 (amplitude and ionization parameter for nebular templates) + 1 (dust attenuation) parameters. 

The SED fitting was performed on \idb. The flux measurements for \ida\ suffer from larger uncertainties, likely due to its location near bright galaxies as well as its smaller magnification factor. The physical quantities of \id\ are thus represented by using the measurements of \idb\ throughout the manuscript. The photometric fluxes used and the derived physical quantities are reported in Tables~\ref{tab:phys} and \ref{tab:flux}. 

{\id\ is characterized with low-stellar mass, \mstelemagnif.} The star formation rate is calculated using the \hb\ flux at the position of AMORE6-B as:
{
\begin{equation}
    {\rm SFR\,[M_\odot\,yr^{-1}]} = 2.1 \times 10^{-42} L_{\rm H\alpha}\,[{\rm erg\,s^{-1}}].
\end{equation}
where the conversion factor was derived from the BPASS model, and} $L_{\mathrm{H}\alpha} = 2.86 \times L_{\mathrm{H}\beta}$ assuming Case B recombination with $T_{\rm e} = 10^{4}$ K and $n_{\rm e} = 10^{2}{\rm cm^{-3}}$ \cite{osterbrock89}. {We note that the \ha-to-\hb\ ratio becomes smaller at a higher density or temperature. The derived SFR, \sfremagnif,} is higher than the one derived using the rest-frame UV luminosity \cite{morishita24} or those averaged over the last 10/100\,Myr of the best-fit star formation history $\mathrm{SFR_{10/100Myr}}$ (Table~\ref{tab:phys}). This supports the presence of a very young burst \cite{caplar19,mcclymont25}.

The best-fit SED is shown in Figure~\ref{fig:sed}. The SED of \id\ is characterized by a very young stellar population. Remarkably, {the UV spectral slope ($f_\lambda \propto \lambda^\beta_{\rm UV}$)
, which is measured with the best-fit SED in the rest-frame wavelength range of $1350 < \lambda/$\AA\,$<3000$, is found to be $\beta_{\rm UV}=$\,\uvbeta. This is considerably blue, likely requiring massive ($>50$--$100\,M_\odot$) stars \cite{eldridge17,schaerer25}. We note that the stellar component of the best-fit model is characterized by an even bluer slope, $\beta_{\rm UV}=$\,\uvbetastel. The slope directly measured with the photometric data points is $\beta_{\rm UV}=$\,\uvbetaphot.}

{
Using the WFSS \hb\ line flux and the continuum flux density of the best-fit SED model, we measured the rest-frame equivalent width to be \ewhb\,$=$\,\ewhberr. This high value is predicted for a very young burst ($<4$\,Myr) with a Pop-III IMF \cite{schaerer02}, consistent with the light-weighted age \tage\ from SED fitting. To cross check the measurements, we also measured \ewhb\ using two photometric filters, F356W (which covers \hb) and F360M (only the continuum red-ward of \hb). We found \ewhb\,$=$\,\ewhbphotom, which is consistent with the measurement above, while the large error originates mainly from the F360M flux error. 

We observe a flux excess in the F444W band from its adjacent bands, which corresponds to the wavelength of \ha. Similarly, using the F444W and F410M (corresponding to the continuum) fluxes, we derived the \ha\ equivalent width, \ewha\,$=$\,\ewhaphotom, which is in agreement with the one for \hb\ after correcting for the Case B \ha-to-\hb\ factor 2.86. The large error originates mainly from the F410M flux error. These equivalent width measurements using photometric fluxes independently confirm the WFSS \hb\ flux measurement, supporting the discussion and interpretation in the present study. 
}

{Following the theoretical framework by \cite{ferrara23}, with an assumed halo mass of $\approx10^8\,M_\odot$, SFR of \id\ is predicted to be $\approx0.00175 (\frac{\epsilon_*}{0.01})$  where ${\epsilon_*}$ is the star formation efficiency. To meet the observed \hb\ SFR, a high ${\epsilon_*}$ is required, $\approx0.4$. Such a high efficiency is achievable in a feedback-free environment, that is, in a pre-SN phase \cite{dekel23,ferrara25}, less than a few\,Myr. This requirement is compatible with \id\ for its light-weighted age (\tage) from SED fitting and the high \ewhb\ measurements reported above. 
}







\captionsetup[figure]{name=Extended Data Figure}
\setcounter{figure}{0}

\begin{figure*}[!htb]
\centering
	\includegraphics[width=0.98\textwidth]{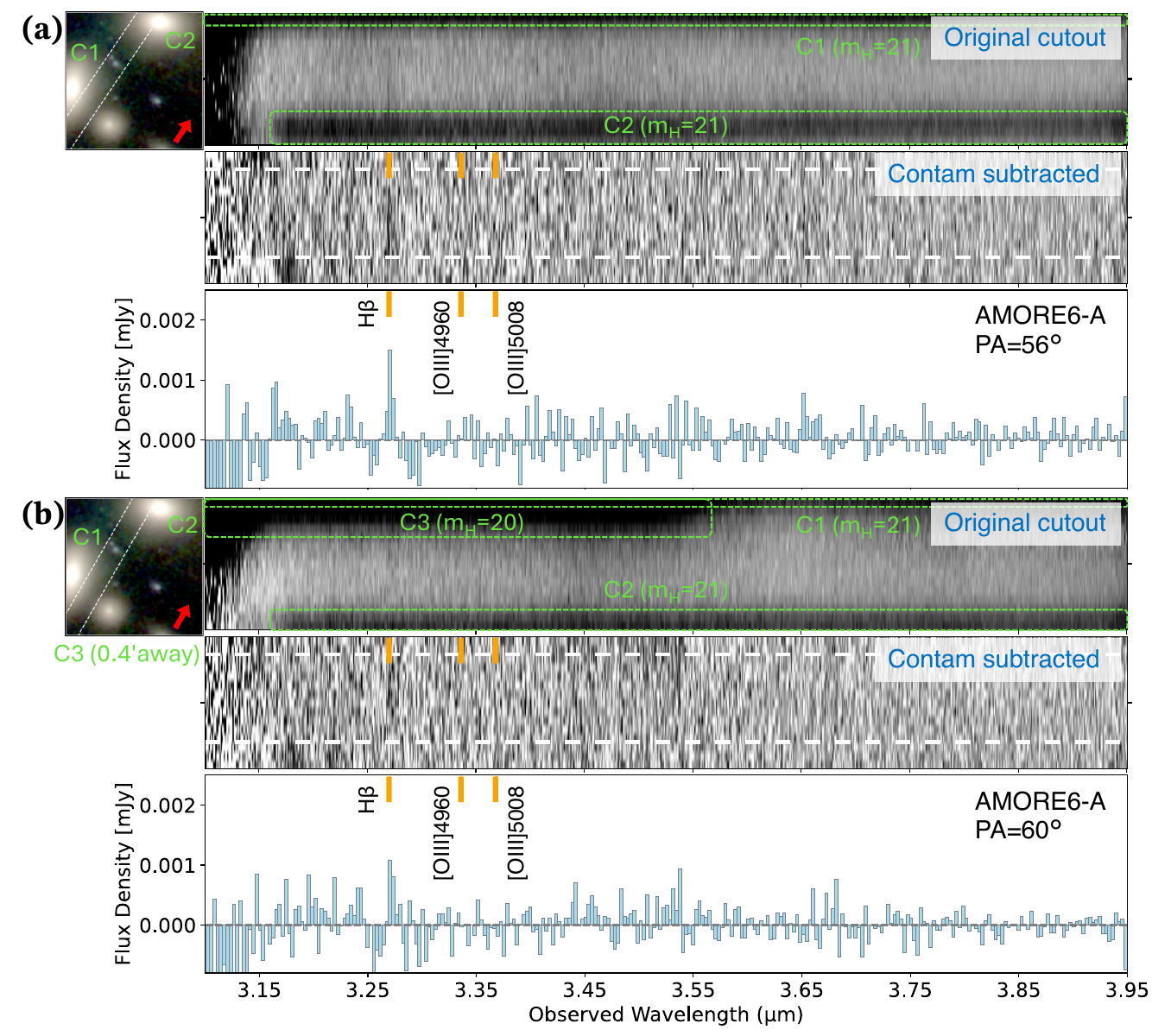}
	\caption{
    {\bf Spectra of \ida\ from individual position angles.}
    Same as Figure~\ref{fig:spectrum} but for spectra of \ida\ from individual position angle data sets, $56^\circ$ ({\bf a}) and $60^\circ$ ({\bf b}). {The spectral trace (white dashed lines; $0.\!''5$ width), along the dispersion direction (red arrow), is shown in the cutout. Spectral contaminants (C1, C2, and C3) are marked in the image stamp and in the spectral cutout image. Note that C3 is $0.\!'44$ away from \id\ to the South-East and is beyond the extent of the cutout stamp.}
    }
\label{fig:spectrum_ind}
\end{figure*}

\begin{figure*}[!htb]
\centering
	\includegraphics[width=0.98\textwidth]{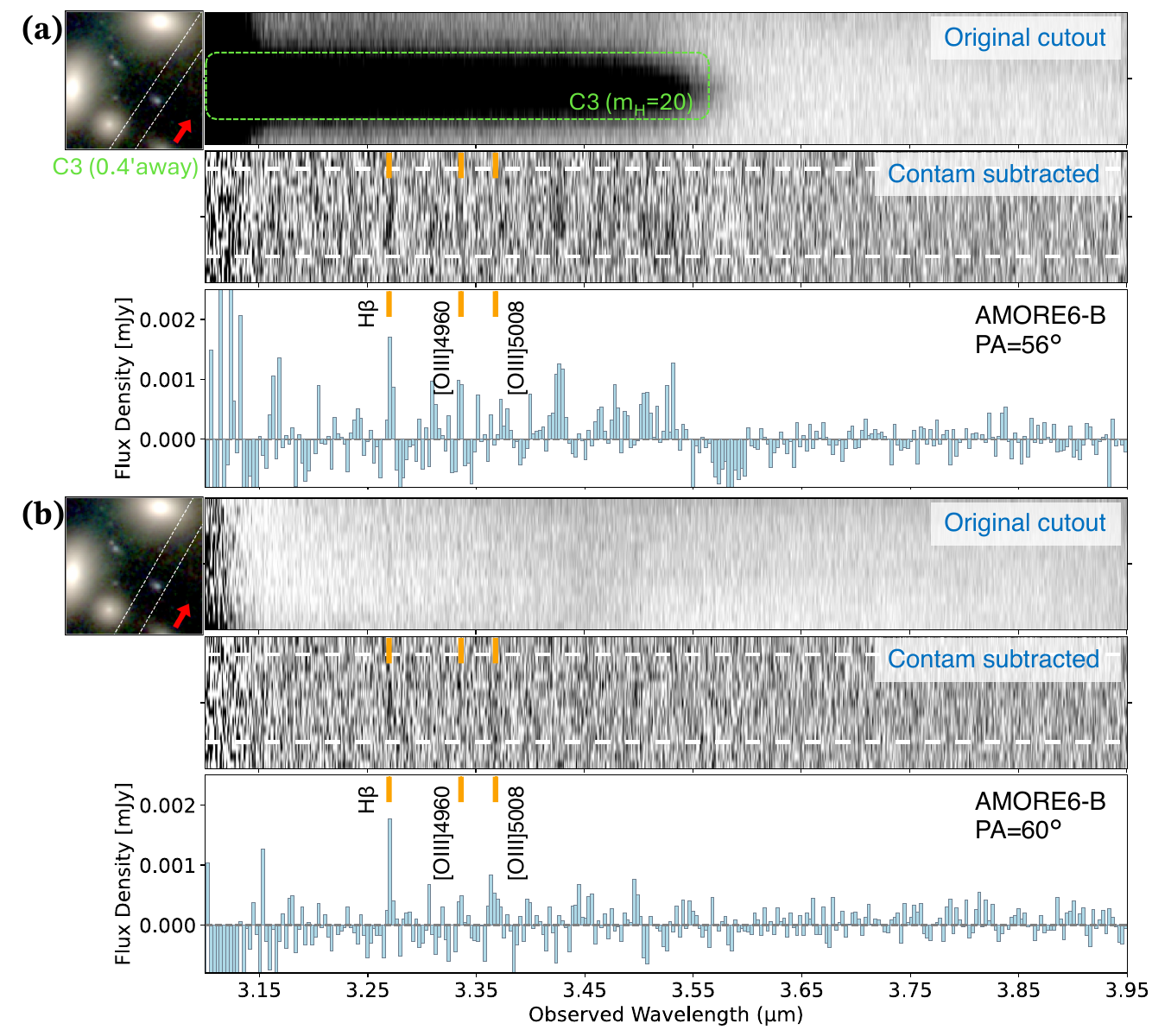}
	\caption{
    {\bf Spectra of \idb\ from individual position angles.}
    Same as Supplementary Data Figure~\ref{fig:spectrum_ind} but for the spectra of \idb.
    }
\label{fig:spectrum_ind_idb}
\end{figure*}

\begin{figure*}[!htb]
\centering
	\includegraphics[width=0.7\textwidth]{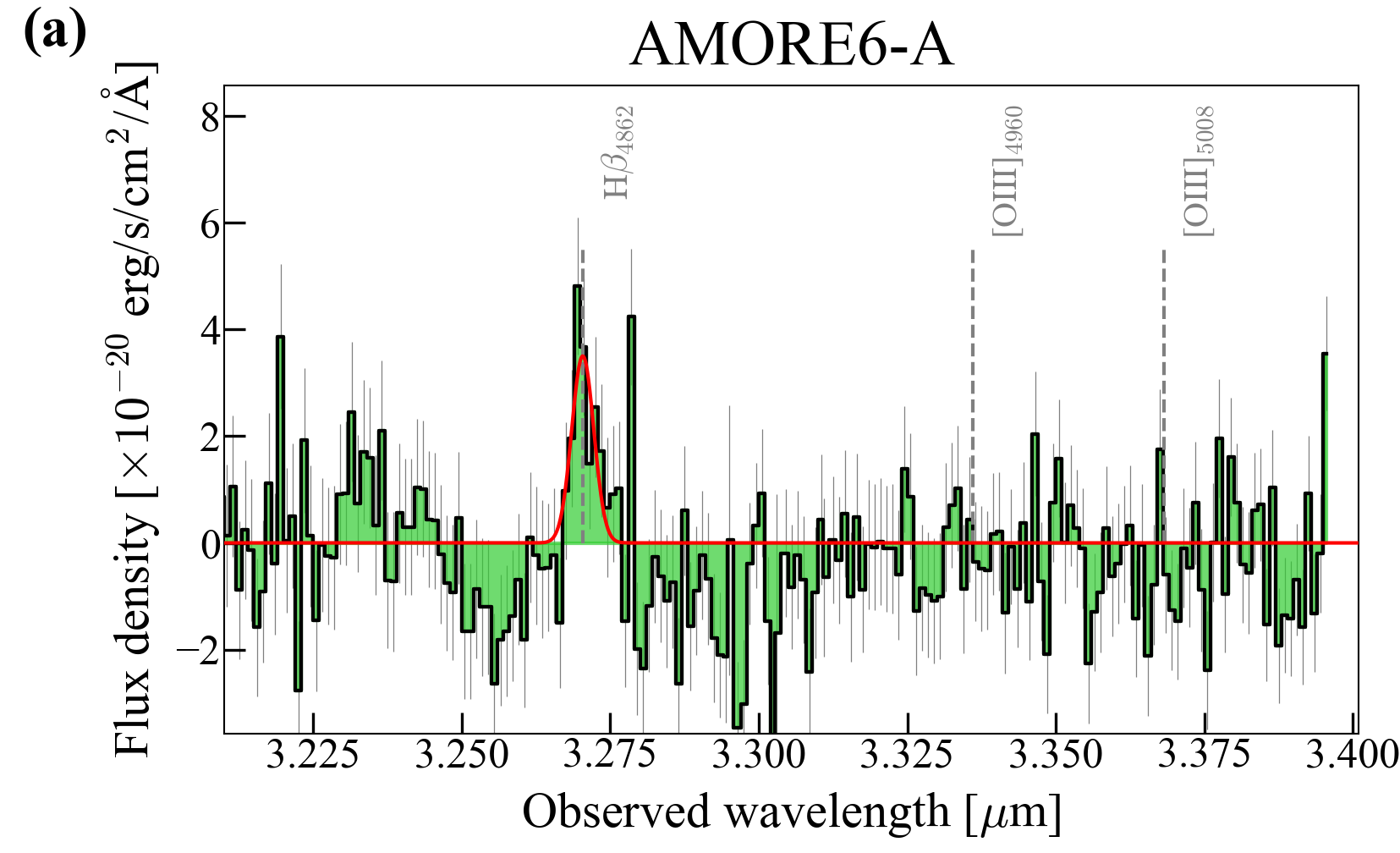}
	\includegraphics[width=0.7\textwidth]{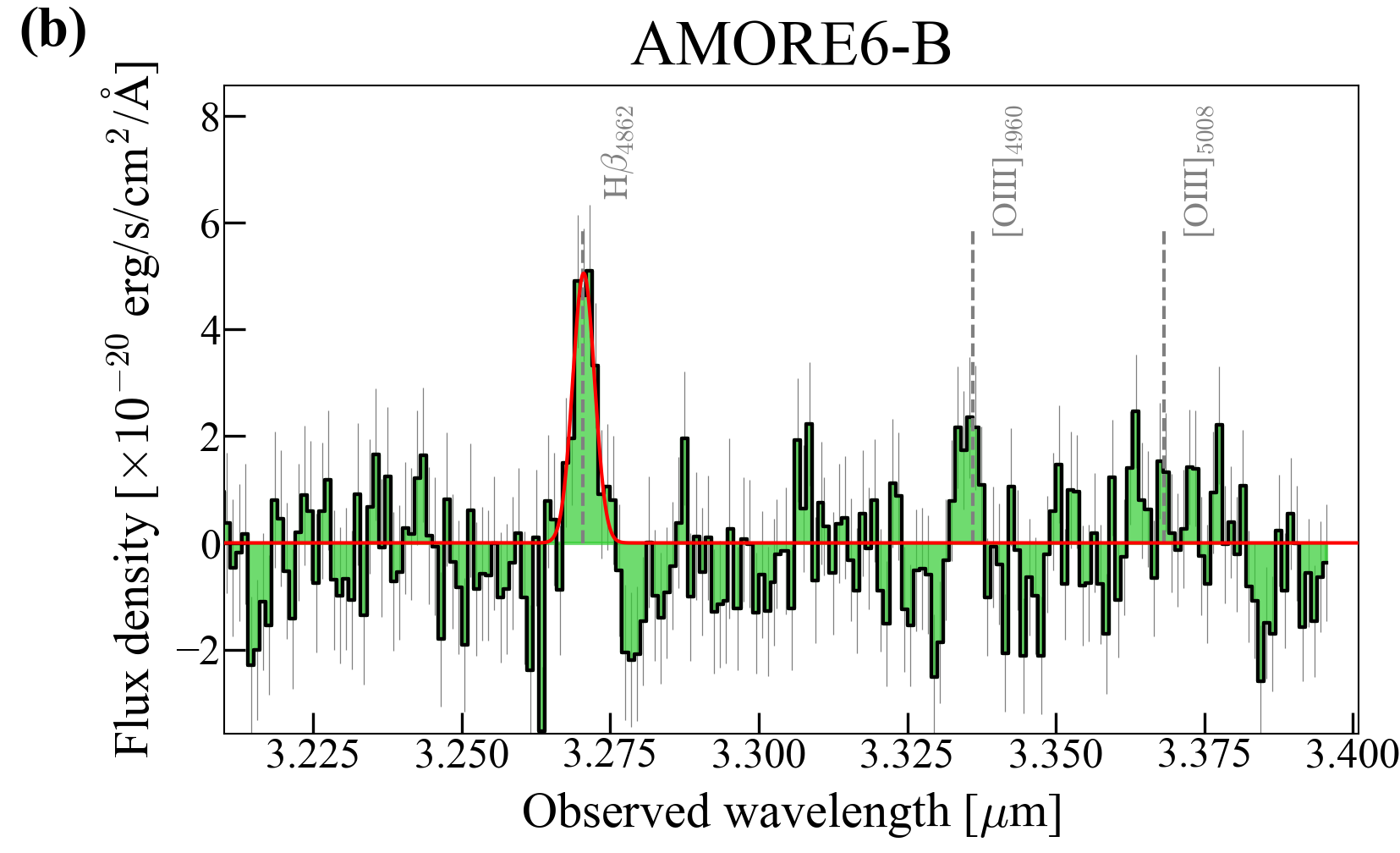}
	\includegraphics[width=0.7\textwidth]{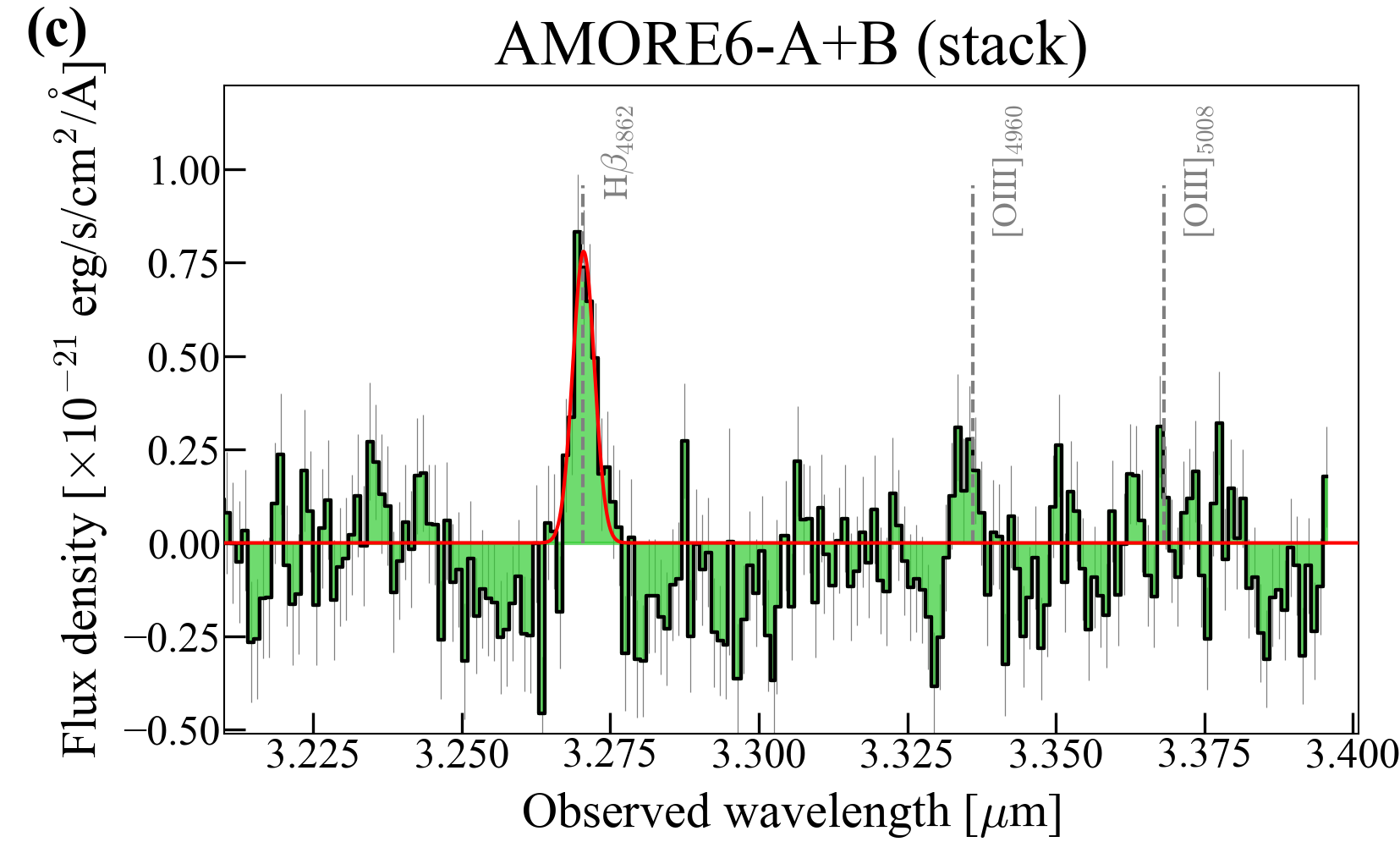}
	\caption{
    {\bf NIRCam WFSS spectra of \id.}
    ({\bf a}) NIRCam WFSS spectrum of \ida\ in the {\bf native spectral element size (10\,\AA}; filled regions with error bars). The best-fit gaussian model is shown (red solid line). The wavelengths of the \oiii-doublet lines are indicated (vertical dashed lines).
    ({\bf b}) Same as the top panel, but for \idb. The potential origin of positive pixels near the wavelengths of $\oiiib$ is discussed in ``NIRCam WFSS Reduction.''  
    ({\bf c}) Same as the top panel, but for the \ida+B stack spectrum. Each spectrum was normalized by magnification before being stacked.
    }
\label{fig:lines}
\end{figure*}

\begin{figure*}[!htb]
\centering
	\includegraphics[width=0.7\textwidth]{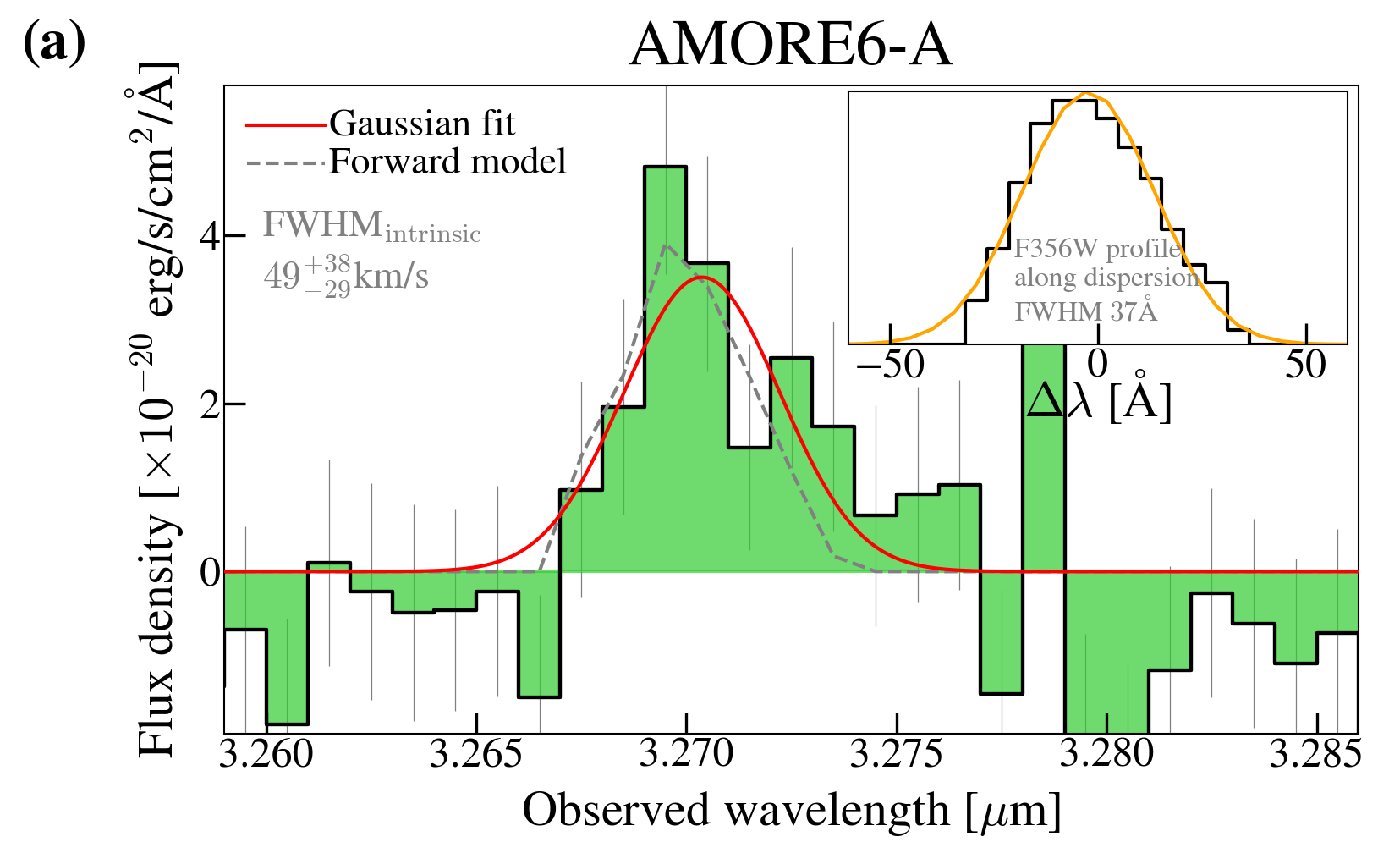}
	\includegraphics[width=0.7\textwidth]{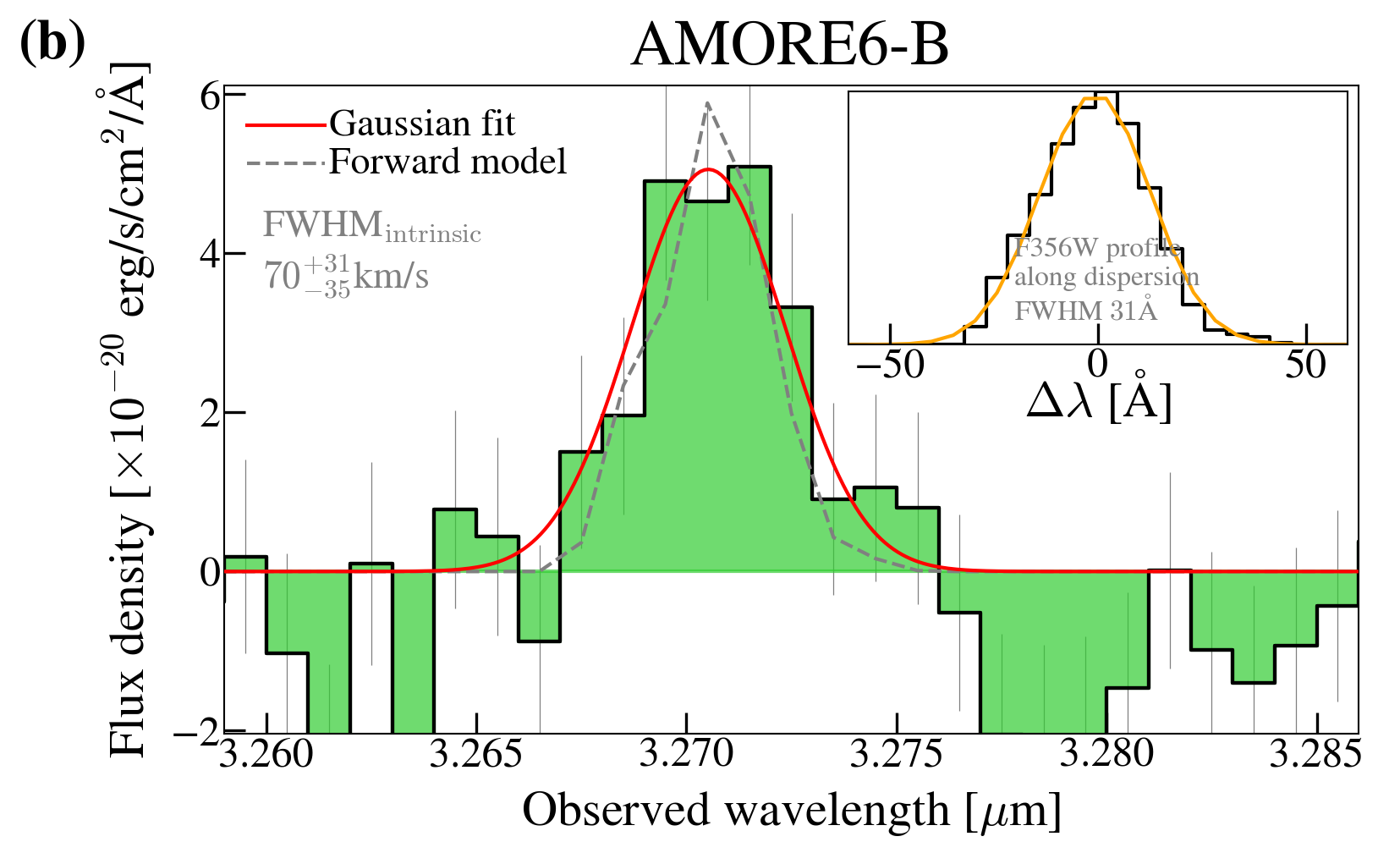}
	\caption{
    {\bf \hb\ lines of \id\ in NIRCam WFSS spectra.}
    {({\bf a}) The same plot as in Figure~\ref{fig:lines}, but zoomed in the \hb\ wavelength. The best-fit model from the forward modeling (see ``\hb\ Emission Line Profile'') is shown (gray dashed line) and compared with the single gaussian profile from Figure~\ref{fig:lines} (red lines). The line spread function is shown in the inset. 
    ({\bf b}) Same as the top panel, but for \idb.
    }
    }
\label{fig:lsf}
\end{figure*}

\begin{figure*}[!htb]
\centering
	\includegraphics[width=1.\textwidth]{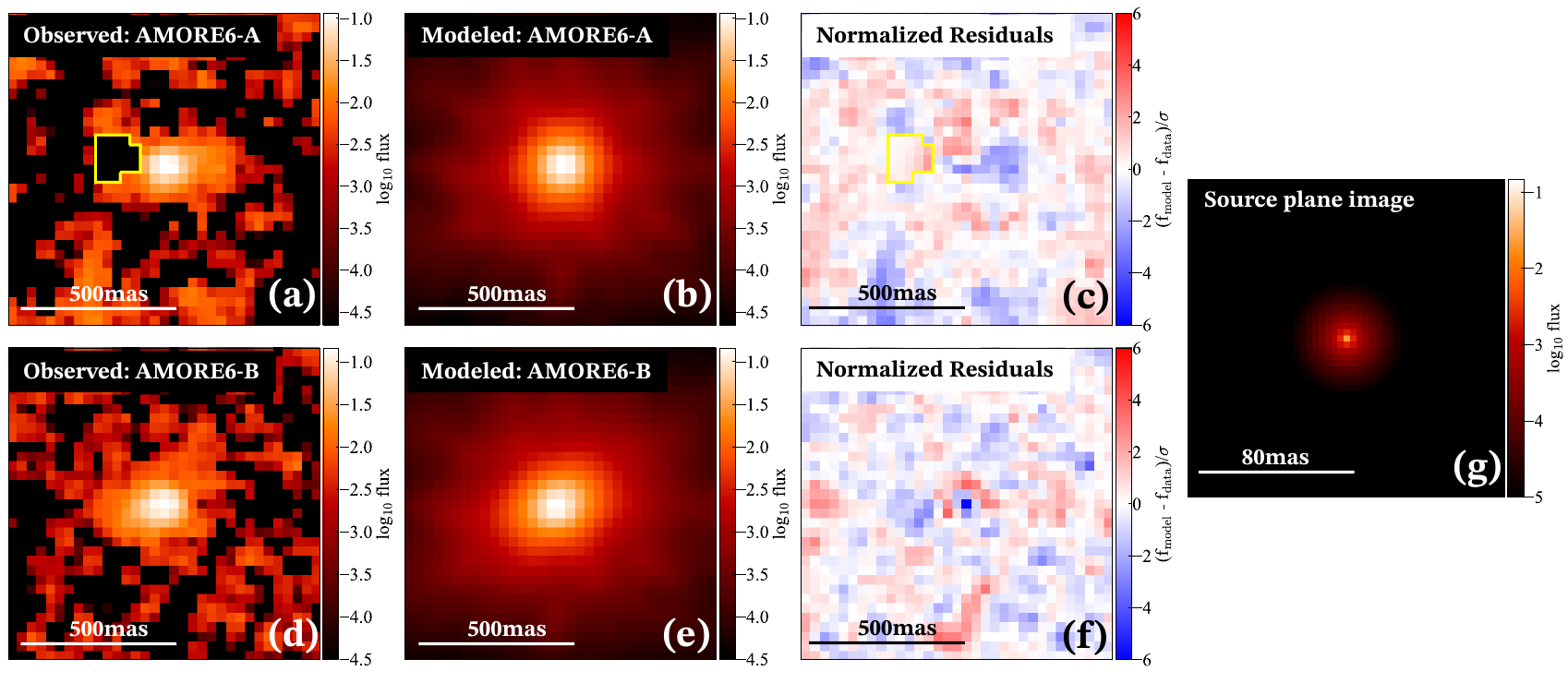}
	\caption{
    {\bf Two-dimensional light profile.}
    {Observed NIRCam F115W image of \ida\ and \idb\ ({\bf a} and {\bf d}, respectively), in a $1''\times1''$ cutout stamp. Forward-modeled two-dimensional light profile ({\bf b}, {\bf e}). Normalized residual map ({\bf c}, {\bf f}). The source plane image is shown in a  $0.\!''16\times0.\!''16$ cutout stamp ({\bf g}). A foreground compact source to the North-East of \ida\ (the region within yellow line boundaries) is masked out during the fitting analysis.}
    }
\label{fig:lenst}
\end{figure*}

\begin{figure*}[!htb]
\centering
	\includegraphics[width=0.99\textwidth]{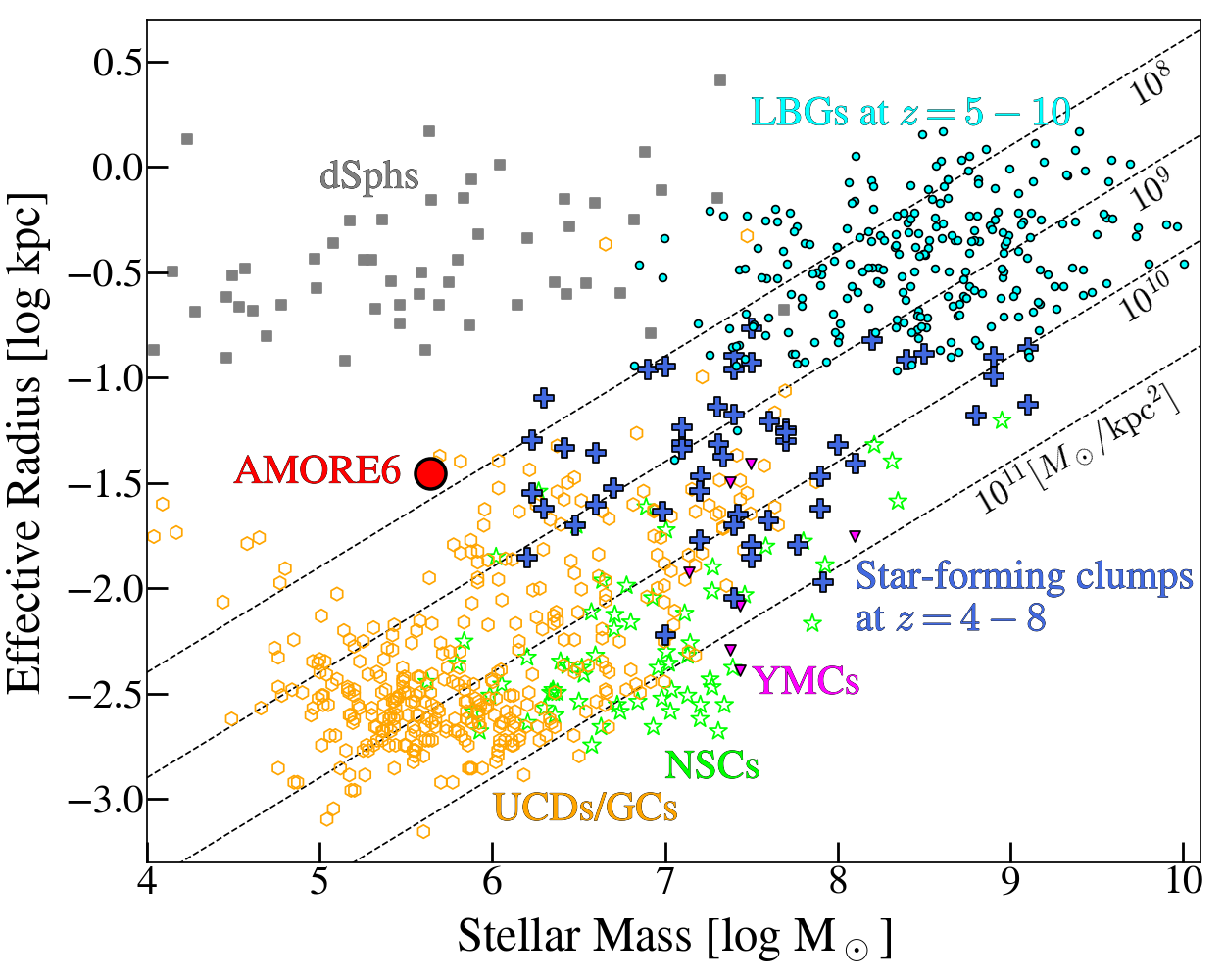}
	\caption{
    {\bf Size-stellar mass distribution.}
    {\id\ (red circle) is compared with various types of objects, including dwarf spheroids (dSphs, gray squares), nuclear star clusters (NSCs, green stars), ultra-compact dwarfs and globular clusters (UCDs/GCs, yellow hexagons), young massive clusters (YMCs, magenta inverted triangles), all at $z\sim0$ from \cite{norris14}. High-$z$ Lyman-break galaxies (cyan circles) \cite{morishita24} and individual star-forming clumps (blue crosses) \cite{messa24a,fujimoto24,morishita25} studies with JWST are also shown.}
    }
\label{fig:sizemass}
\end{figure*}

\begin{figure*}[!htb]
\centering
	\includegraphics[width=0.9\textwidth]{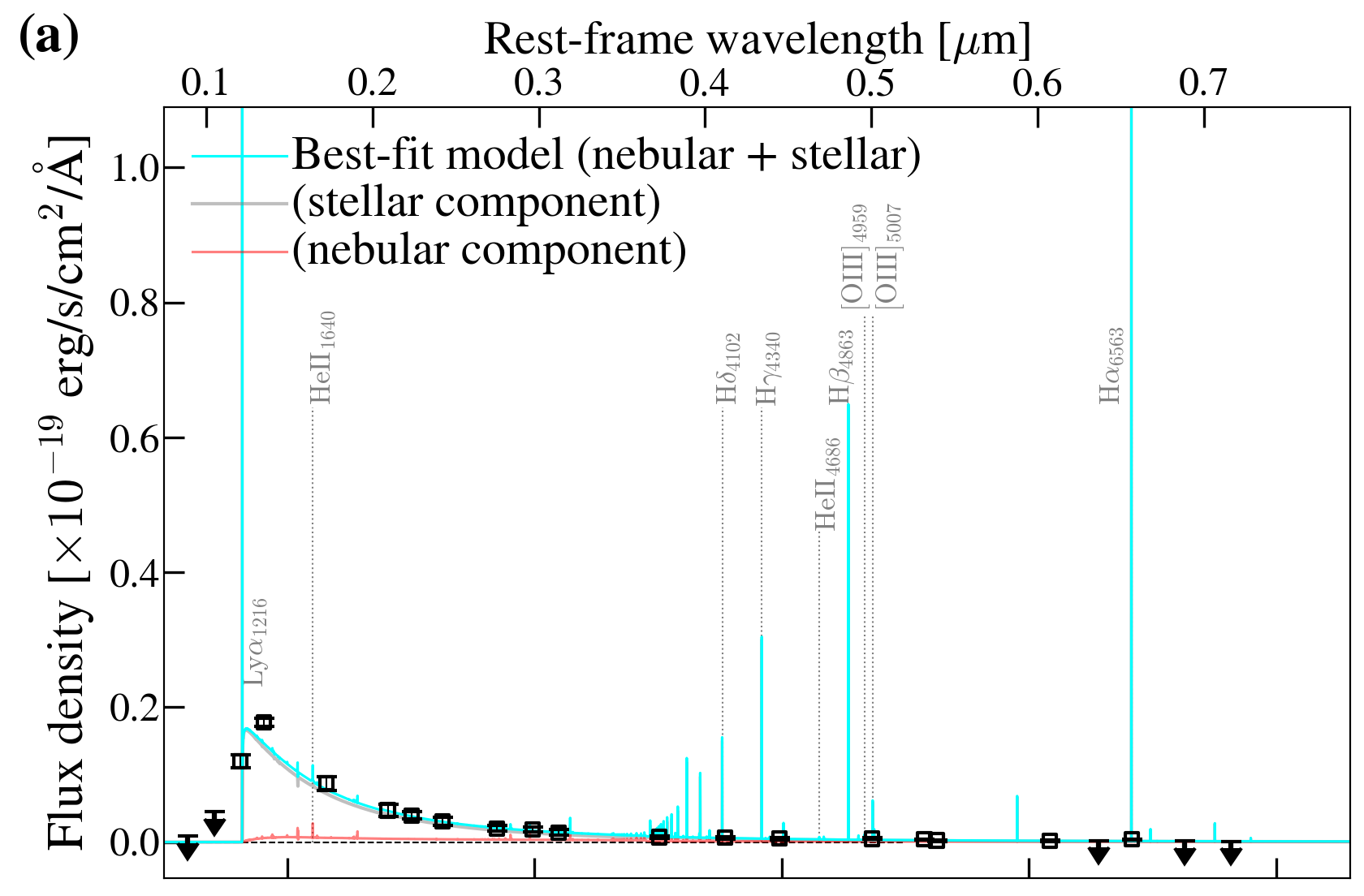}
	\includegraphics[width=0.9\textwidth]{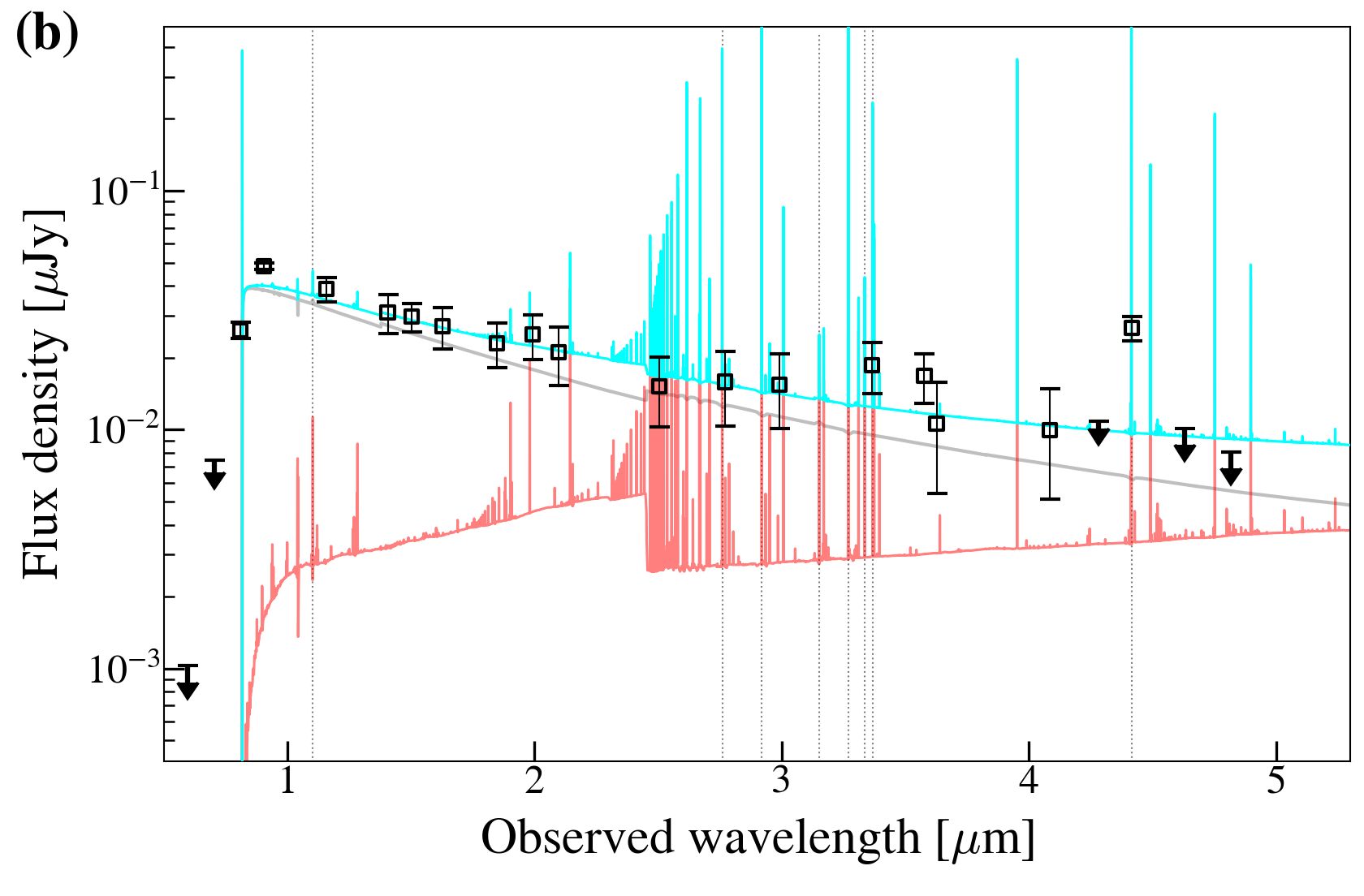}
	\caption{
    {\bf Spectral energy distribution of \id.}
    ({\bf a})
    The spectral energy distribution is modeled using the photometry of \idb, in $f_\lambda$. Observed photometric fluxes are shown with 1\,$\sigma$ errors/upper limits (squares/arrows). The best-fit full model spectrum (nebular+stellar; cyan line), stellar component (gray line), and nebular component (red line) are shown.
    ({\bf b}) Same as the top panel but in $f_\nu$. 
    }
\label{fig:sed}
\end{figure*}


\clearpage

\captionsetup[table]{name=Supplementary Data Table}

\begin{table}[h!]
\centering
\caption{
{\bf Observed properties of \ida\ and \idb.} $1\,\sigma$ errors are quoted. For measurements with non-detection involved, \qsigma\,$\sigma$ upper limits are quoted. 
$\dagger$: The stacked spectrum \id-A+B was normalized by the magnification factor before being stacked. 
{$\ddagger$: Measured \hb\ line width from single-component gaussian fitting. Note that the observed line width is dominated by convolution effect of WFSS (see ``NIRCam Wide Field Slitless Spectroscopy Data Reduction'').}
}
\begin{tabular}{ccccc}
\toprule
{Quantity} & {Unit} & {\ida} & {\idb} & {\id-A+B}$^{\dagger}$\\
\midrule
R.A. & degree & $3.584723$ & $3.584413$ & --\\
Decl. & degree & $-30.40316$ & $-30.40340$ & --\\
$\mu$ &  & $39.3_{-3.5}^{+3.7}$ & $77.7_{-5.9}^{+8.4}$ & $1$\\
$z(\mathrm{H\beta})$ &  & $5.7253_{-0.0010}^{+0.0011}$ & $5.7253_{-0.0005}^{+0.0005}$ & $5.7253_{-0.0005}^{+0.0005}$\\
$f({\rm H\beta}_{4862})$ & $10^{-19}$\,erg/s/cm$^2$ & $15.73 \pm 3.82$ & $22.64 \pm 3.65$ & $0.35 \pm 0.05$\\
$f({\rm [O\textsc{iii}]}_{4960})$ & $10^{-19}$\,erg/s/cm$^2$ & $<9.90$ & $<9.57$ & $<0.12$\\
$f({\rm [O\textsc{iii}]}_{5008})$ & $10^{-19}$\,erg/s/cm$^2$ & $<9.48$ & $<9.82$ & $<0.12$\\
${\rm FWHM(H\beta)}^{\ddagger}$ & \AA & $43.5\pm11.3$ & $42.4\pm6.1$ & $42.8\pm5.4$\\
${\rm EW_0(H\beta)}$ & \AA\ & $1289_{-386}^{+456}$ & $939_{-210}^{+248}$ & $1122_{-220}^{+260}$\\
R3 &  & $<0.60$ & $<0.43$ & $<0.33$\\
$12 + \mathrm{\log (O/H)}$ &  & $<6.24$ & $<6.09$ & $<5.97$\\
$f({\rm Ly\alpha}_{1216})/f({\rm H\beta}_{4862})$ &  & -- & -- & $14.7\pm2.0$\\
\bottomrule
\end{tabular}
\label{tab:lines}
\end{table}

\begin{table}[h!]
\centering
\caption{
{\bf Physical properties of \id.} SED Measurements are represented by AMORE6-B and corrected for magnification using $\mu=$\,\magniferr\ \cite{bergamini23b}. {The effective radius is estimated with {\tt lenstruction}. Where relevant, the quoted uncertainties include the magnification error.}
}
\begin{tabular}{ccc}
\toprule
{Property} & {Unit} & {Measurement}\\
\midrule
Stellar mass $M_*$ & $\log M_\odot$ & $5.64_{-0.08}^{+0.18}+\log(78/\mu)$\\
Absolute UV magnitude $M_\mathrm{UV}$ & mag & $-14.52_{-0.08}^{+0.07}-2.5\log(78/\mu)$\\
UV slope $\beta_{\mathrm{UV}}$ &  & $-2.77_{-0.09}^{+0.07}$\\
\hb-based Star formation rate $\mathrm{SFR_{H\beta}}$ & $\log M_\odot {\rm yr^{-1}}$ & $-1.18_{-0.08}^{+0.06}+\log(78/\mu)$\\
UV-based Star formation rate $\mathrm{SFR_{UV}}$ & $\log M_\odot {\rm yr^{-1}}$ & $-1.73_{-0.09}^{+0.07}+\log(78/\mu)$\\
SFH-based Star formation rate $\mathrm{SFR_{10Myr}}$ & $\log M_\odot {\rm yr^{-1}}$ & $-1.47_{-0.12}^{+0.10}+\log(78/\mu)$\\
SFH-based Star formation rate $\mathrm{SFR_{100Myr}}$ & $\log M_\odot {\rm yr^{-1}}$ & $-2.42_{-0.09}^{+0.16}+\log(78/\mu)$\\
Source plane effective radius $R_\mathrm{e}$ & pc & $36.0_{-7.5}^{+9.4}\times (78/\mu)^{1/2}$\\
Stellar mass surface density $\Sigma_*$ & $\log M_\odot {\rm pc^2}$ & $2.26_{-0.08}^{+0.18}$\\
Star formation rate surface density $\Sigma_{\rm SFR}$ & $\log M_\odot {\rm yr^{-1} kpc^2}$ & $2.16_{-0.08}^{+0.06}$\\
Light-weighted age $t$ & ${\rm Myr}$ & $1.5_{-1.0}^{+1.2}$\\
\bottomrule
\end{tabular}
\label{tab:phys}
\end{table}

\begin{table}[h!]
\centering
\caption{
{\bf Measured fluxes of \idb.} Fluxes detected at S/N$>2$ are associated with $1\,\sigma$ uncertainties; \qsigma\,$\sigma$ upper limits are shown for non-detections. Flux measurements are not corrected for magnification. 
}
\begin{tabular}{cc}
\toprule
{Filter} & {Flux}\\
{} & {nJy}\\
\midrule
HST\_ACS\_F435W & $<5.0$\\
HST\_ACS\_F606W & $<4.6$\\
HST\_ACS\_F814W & $26.1\pm2.1$\\
JWST\_NIRCAM\_F070W & $<24.0$\\
JWST\_NIRCAM\_F090W & $48.4\pm1.5$\\
JWST\_NIRCAM\_F115W & $38.8\pm4.6$\\
JWST\_NIRCAM\_F140M & $31.0\pm5.6$\\
JWST\_NIRCAM\_F150W & $29.7\pm4.0$\\
JWST\_NIRCAM\_F162M & $27.1\pm5.4$\\
JWST\_NIRCAM\_F182M & $23.1\pm4.9$\\
JWST\_NIRCAM\_F200W & $25.0\pm5.3$\\
JWST\_NIRCAM\_F210M & $21.1\pm5.9$\\
JWST\_NIRCAM\_F250M & $15.1\pm4.9$\\
JWST\_NIRCAM\_F277W & $15.9\pm5.5$\\
JWST\_NIRCAM\_F300M & $15.4\pm5.3$\\
JWST\_NIRCAM\_F335M & $18.6\pm4.5$\\
JWST\_NIRCAM\_F356W & $16.8\pm3.9$\\
JWST\_NIRCAM\_F360M & $10.6\pm5.2$\\
JWST\_NIRCAM\_F410M & $10.0\pm4.8$\\
JWST\_NIRCAM\_F430M & $<22.0$\\
JWST\_NIRCAM\_F444W & $26.6\pm3.1$\\
JWST\_NIRCAM\_F460M & $<44.1$\\
JWST\_NIRCAM\_F480M & $<34.9$\\
\bottomrule
\end{tabular}
\label{tab:flux}
\end{table}







\end{document}

%% file: authors.tex
\author{
	Takahiro~Morishita$^{1,2\ast}$,
	Zhaoran~Liu$^{2, 3}$,
	Massimo~Stiavelli$^{4}$,
	Tommaso~Treu$^{5}$,\\
	Pietro~Bergamini$^{6,7}$,
	Yechi~Zhang$^{1}$\and
	\small$^{1}$IPAC, California Institute of Technology, Pasadena, CA 91125, USA.\and
    \small$^{2}$Astronomical Institute, Tohoku University, 6-3, Aramaki, Aoba-ku, Sendai, Miyagi 980-8578, Japan.\and
	\small$^{3}$MIT Kavli Institute for Astrophysics and Space Research, 70 Vassar Street, Cambridge, MA 02139, USA.\and
	\small$^{4}$Space Telescope Science Institute, Baltimore, MD 21218, USA.\and
	\small$^{5}$Department of Physics and Astronomy, University of California, Los Angeles, CA 90095, USA.\and
	\small$^{6}$Dipartimento di Fisica, Università degli Studi di Milano, Milano I-20133, Italy.\and
	\small$^{7}$INAF - OAS, Osservatorio di Astrofisica e Scienza dello Spazio di Bologna, Bologna I-40129, Italy.\and
	\small$^\ast$Corresponding author. Email: takahiro@ipac.caltech.edu.
}

%% file: materials.tex
\subsection*{Supplementary materials}
Materials and Methods\\
Supplementary Text\\
Figs. S1 to S7\\
Tables S1 to S3\\
References ({\it 51}-{\it 90})\\